\documentclass{iopjournal}
\usepackage{graphicx}

\usepackage{url}
\usepackage{subcaption}
\usepackage[numbers,sort&compress]{natbib}

\usepackage[utf8]{inputenc}
\usepackage[T1]{fontenc}
\usepackage{amsmath}

\hypersetup{colorlinks,linkcolor=[rgb]{0.2,0.2,0.75},citecolor=[rgb]{0.3,0.5,0.5},urlcolor=[rgb]{0.4,0.4,0.6},breaklinks}

\begin{document}

\articletype{Paper}

\title{AXUV synthetic diagnostic for ASDEX Upgrade and its application for SPI simulations}

\author{Ferenc Lengyel$^{1,2,*}$\orcid{0009-0000-1003-2530}, Weikang Tang$^{3,*}$\orcid{0000-0002-8406-8349}, Matthias Hölzl$^{3,4}$\orcid{0000-0001-7921-9176}, Matthias Bernert$^3$\orcid{0000-0003-1131-0867}, Matěj Tomeš$^5$\orcid{0000-0001-9477-8398}, Peter Halldestam$^3$\orcid{0000-0001-6471-2304}, Paul Heinrich$^3$\orcid{0000-0003-1823-5257}, Gergely Papp$^3$\orcid{0000-0003-0694-5446}, Stefan Jachmich$^6$, Umar Sheikh$^7$, Mathias Dibon$^6$\orcid{0000-0003-2694-6956}, Pascal de Marné$^3$, Jörg Hobirk$^3$\orcid{0000-0001-6605-0068}, Thomas Eberl$^3$, Gergő I. Pokol$^{1,2}$\orcid{0000-0003-1473-0736}, the ASDEX Upgrade Team$^a$ and the EUROfusion Tokamak Exploitation Team$^b$}

\affil{$^1$Department of Nuclear Techniques, Budapest University of Technology and Economics, Budapest, 1111, Hungary}
\affil{$^2$Institute for Atomic Energy Research, HUN-REN Centre for Energy Research, Budapest, 1121, Hungary}
\affil{$^3$Max Planck Institute for Plasma Physics, 85748 Garching, Germany}
\affil{$^4$Department of Physics and Astronomy, Chalmers University of Technology, Göteborg, SE-41296, Sweden}
\affil{$^5$Institute of Plasma Physics, Czech Academy of Sciences, 182 11 Praha 8, Czech Republic}
\affil{$^6$ITER Organization, Route de Vinon sur Verdon, 13067 Saint Paul Lez Durance, France}
\affil{$^7$Swiss Plasma Center, Ecole Polytechnique Fédérale de Lausanne, Lausanne CH-1015, Switzerland}
\affil{$^a$See the author list of T. Pütterich et al. 2026 \textit{Nucl. Fusion} \textbf{66} 116002}
\affil{$^b$See the author list of N. Vianello et al. 2026 \textit{Nucl. Fusion} \textbf{66} 116010}
\affil{$^*$Authors to whom any correspondence should be addressed.}

\email{lengyel.ferenc@ek.hun-ren.hu, weikang.tang@ipp.mpg.de}


\begin{abstract}

We introduce an Absolute eXtended UltraViolet (AXUV) diode-based camera forward-modelling tool to support the validation of mitigated disruption simulations and the interpretation of experimental phenomena, with applications to the ASDEX Upgrade (AUG) tokamak.
AXUV diodes measure electromagnetic radiation across a wide spectral range with a significantly higher time resolution (${\sim}\mu\mathrm{s}$) than foil bolometers (${\sim}\mathrm{ms}$), albeit with a non-uniform spectral responsivity.
AXUV is suitable for examining fast phenomena, such as shattered pellet injection (SPI), where the radiation localisation and radiated power provide information on the deposition of pellet material.
Due to the characteristics and degradation of AXUV diodes, absolute power measurements are subject to large systematic uncertainties, especially when the spectra are time-varying, as in e.g. mixed Ne/D$_2$ SPI experiments.
These challenges motivated the development of a synthetic diagnostic within the Cherab-Raysect optical modelling framework, which is applied here to four AXUV cameras in two poloidal cross-sections of AUG.
The synthetic diagnostic provides a means to understand how the diodes measure radiation under SPI conditions and to connect first-principles plasma simulations with experimental measurements.
The details of the synthetic diagnostic are presented, and the capabilities are illustrated with applications to AUG SPI simulations performed in JOREK\@.
The synthetic signals generated from these simulations are compared with experimental measurements from the 2022 SPI campaign and show qualitatively similar features in many respects.
Particularly good agreement was found in the time evolution of the studied high Ne-content (10\%) pellet, whereas a different, low Ne-content (0.17\%) case exhibited more pronounced differences, likely due to the absence of background impurities in the underlying SPI simulations.

\end{abstract}

\section{Introduction}

Tokamaks are prone to disruptions that threaten their operation, and effective mitigation is an important task for reactor-scale machines~\citep{bandyopadhyay_mhd_2025,hender_chapter_2007}.
Shattered pellet injection (SPI) has been chosen as the primary method for the ITER Disruption Mitigation System~\citep{lehnen_disruptions_2015,loarte_new_2025, Barabaschi_2026} with plans for pure protium and mixed protium-neon pellets~\citep{lee_peridynamic_2024,hu_plasmoid_2024}.
Experimental validation of SPI and related mitigation techniques is currently ongoing on multiple tokamak devices~\cite{jachmich_shattered_2021, Bodner_2025}, including ASDEX Upgrade (AUG)~\citep{dibon_design_2023, Heinrich_2025_PhD, Schwarz_2023, Heinrich_2024_SPI_Lab, Schwarz_2024_PhD, Vallhagen_2025, Heinrich_2026_disr_evo, halldestam_reduced_2025, tang_non-linear_2025, tang_quantitative_2026, patel_modelling_2023}.
In order to support the development of a physics basis for disruption mitigation using SPI, a versatile SPI system~\citep{dibon_design_2023} was installed in the ASDEX Upgrade tokamak along with new Absolute eXtended UltraViolet (AXUV) cameras in 2022 to complement the existing set~\citep{bernert_application_2014}.
AXUV diodes detect the significant line radiation expected mainly from the impurity in the pellet material, offer higher time and spatial resolution than foil bolometers, and can aid in interpreting the results of the SPI experiments.
Previous research had already utilised the AXUV diodes on AUG, alongside the foil bolometers, for radiation localisation and asymmetry studies to determine the disruption mitigation efficiency of Massive Gas Injection (MGI)~\citep{sheikh_disruption_2020}.

During AXUV data analysis of SPI disruptions, attempts were made to determine the location of the radiation front in the poloidal cross-section, as well as the radiated power during the SPI-induced disruptions~\citep{patel_modelling_2023,lengyel_plasma_2024,heinrich_radiated_2025}.
It has been found, however, that utilising the AXUV measurements for said purposes is challenging, partly owing to the geometry of the lines of sight (LOSs), since at the SPI injection location, the cameras in the sector have antiparallel LOSs.
Additionally, the non-uniform spectral response function of the diodes degrades during plasma operation~\citep{bernert_application_2014}.
Furthermore, the diodes are calibrated against standard plasma discharges, but this calibration is invalidated whenever the injected pellet contains neon.
Neon radiates throughout the sensitive spectral region of the diodes, which makes the absolute power determination susceptible to systematic error.
This affected more than half of the campaign, since 102 of the 195 successful pellet launches contained neon.
A good characterisation of the radiation dynamics during mixed Ne/D$_2$ SPI is nonetheless key to understanding the thermal quench (TQ) onset conditions, and the effectiveness of thermal and electromagnetic load mitigation.
Additionally, the impurity content of the plasma prior the injection would make this determination quite challenging in any other discharge.
Motivated by the aforementioned uncertainties arising from the AXUV diagnostic, where the emitted radiation spectrum of the plasmas is often not precisely known, this paper presents a 3D synthetic diagnostic for AUG AXUV cameras to assist in interpreting modelling results relative to experimental data.

The main tool used for the modelling presented in this paper is Cherab, an open-source Python library for optical diagnostics modelling based on spectroscopic plasma emission~\citep{carr_towards_2017,carr_cherabcore_2023}.
It is built around the Raysect library, which provides a robust ray-tracing framework~\citep{meakins_raysectsource_2023}.
As Raysect was primarily developed for visible light, a methodology was developed to handle the wide spectral range required for AXUV diodes.
Cherab's ability to handle arbitrary geometries and spectral-dependent sensitivity makes it well-suited to addressing the challenges encountered in the modelling of AXUV systems.

JOREK is a non-linear extended magnetohydrodynamics (MHD) code that models the 3D evolution of tokamak disruptions, including shattered pellet injection and the resulting thermal and current quench dynamics, using increasingly complex physics models~\citep{hoelzl_jorek_2021,czarny_bezier_2008,huysmans_mhd_2007}.
The resulting time- and space-resolved plasma state can be post-processed into synthetic diagnostic signals, enabling direct comparison with experimental measurements, such as those from AXUV bolometry presented in the paper.

The rest of the paper is organised as follows. Section~\ref{sec:AUG-AXUV} presents technical details about the AXUV diodes on AUG, their spectral responsivity, and the visualisation of the lines of sight.
Section~\ref{sec:single-LOS} presents the single-line-of-sight model developed for spatially 1D plasma simulations, which has been used for estimating the time evolution of the detection efficiency of the diodes during an SPI experiment.
Section~\ref{sec:3D-synth-diag} presents the developed 3D synthetic diagnostic, which is more suitable for use with higher-fidelity simulations, such as those from the JOREK code~\citep{tang_non-linear_2025,tang_quantitative_2026,hoelzl_jorek_2021}.
Comparisons of post-processed JOREK modelling results with AXUV signals from experiments are examined in subsection~\ref{subsec:synth-exp-comp}.

\section{AXUV cameras at ASDEX Upgrade}\label{sec:AUG-AXUV}

AXUV diodes detect electromagnetic radiation across a wide spectral range from the visible to the soft X-ray (SXR) range~\citep{tal_measurement_2015,bernert_application_2014,veres_fast_2009,boivin_high_1999}.
The entrance window of these diodes is only 8~nm thick, but even this very thin glass layer absorbs a significant portion of the UV radiation, contributing to a non-uniform spectral sensitivity~\citep{tal_measurement_2015,canfield_stability_1989,korde_quantum_1987}.
Additionally, the spectral responsivity of the diodes degrades due to extreme UV (XUV) photons that break up covalent bonds in the crystalline structure of the entrance window, and due to vacuum UV (VUV) radiation emitted during normal plasma operation~\citep{tal_measurement_2015,bernert_application_2014}.

\begin{figure}
  \centering
  \includegraphics[width=0.8\textwidth]{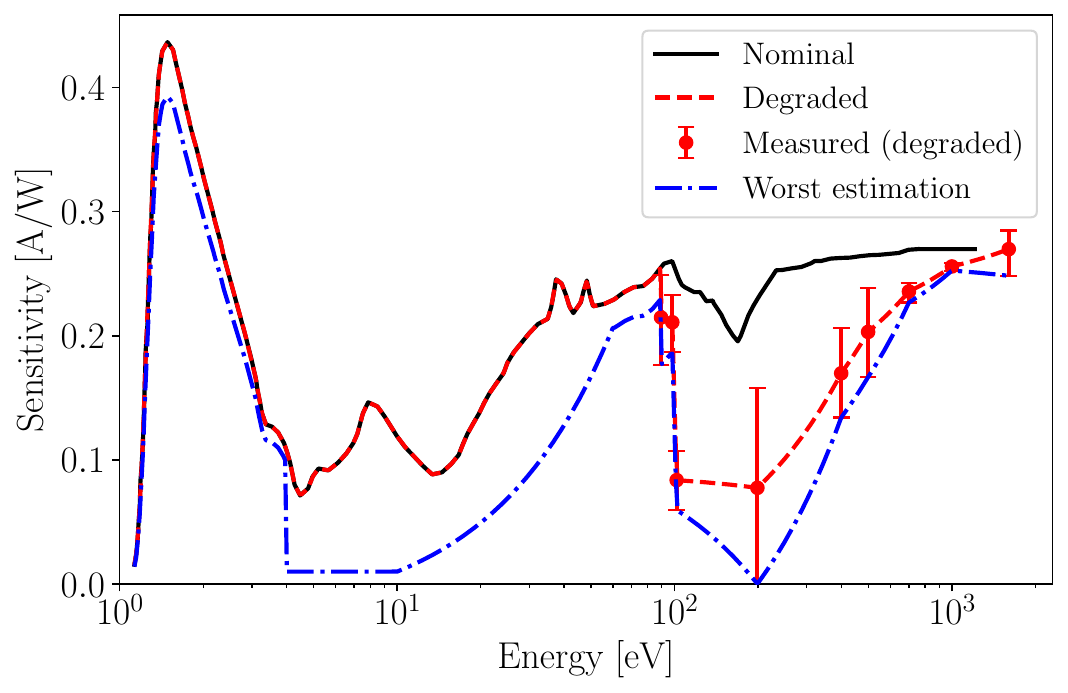}
  \caption{The spectral response of the AXUV diodes as a function of incident photon energy.
  The black line shows the manufacturer-specified nominal responsivity, with the red markers showing the degraded responsivity based on~\citep{bernert_application_2014}, determined after one campaign from measurements performed on several diodes.
  The red dashed line shows the estimated degraded spectral responsivity that was used in the paper for the synthetic diagnostic. The blue dash-dotted line represents a worst-case estimate of the responsivity.}\label{fig:spectral_response}
\end{figure}

The nominal spectral responsivity, according to the manufacturer's specification, for all the AXUV diodes used at AUG is shown in figure~\ref{fig:spectral_response} as the black line.
The red dashed line shows an average degraded responsivity of the AXUV diodes after 4800 seconds of plasma operation~\citep{bernert_application_2014}.
There is significant degradation compared to the new diodes in the 100--400 eV range, but the energy range in which measurements could be performed to determine the degradation was limited to 50--1600 eV, with exceptionally large errors near 50 eV.
Therefore, at lower energies, the nominal sensitivity values were used for the degraded function.
It is also important to note that sensitivity below 100eV also undergoes degradation, meaning that the degraded sensitivity shown in figure~\ref{fig:spectral_response} is an underestimation of the damage these diodes undergo.
The VUV region (4--10 eV) is particularly susceptible to degradation, and sensitivity is reported to be significantly reduced in this range~\citep{bernert_application_2014}.
The AXUV cameras in sector 16, where the shattered pellet injection takes place, were installed specifically for the 2021/22 campaign, with new diodes installed in the sector 5 cameras.
The pre-existing cameras had their lines of sight adjusted to better match those of the new cameras in sector 16.
In conclusion, the exact state of the diode sensitivity functions is unknown in practice, and the radiation reaching the specific diodes may differ, potentially resulting in different degrees of degradation even for neighbouring diodes.
The degraded sensitivity shown in the plot was chosen for this article as an approximation, but we note that this is subject to a large, order-unity uncertainty.

\begin{figure}
  \centering
  \includegraphics[width=0.5\textwidth]{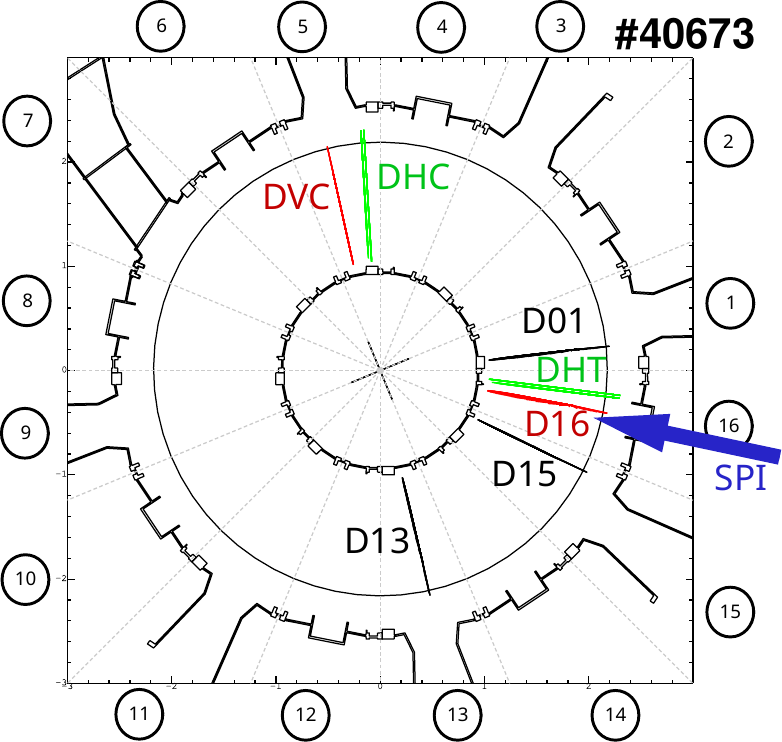}
  \caption{The toroidal cross-section of the ASDEX Upgrade tokamak with some of the AXUV lines of sight, and the SPI injection location.
  }\label{fig:axuv_toroidal}
\end{figure}

There are currently 14 cameras with AXUV diode arrays installed at ASDEX Upgrade~\citep{bernert_application_2014}.
Two types of cameras were considered for this paper, both consisting of 3 arrays of 16 diodes, for a total of 48 diodes.
The vertical-type camera is located at the upper part of the low-field side (LFS), and such cameras were present in five toroidal sectors during the 2021/22 campaign.
The vertical-type cameras used in this paper are the `D16' (sector 16 of AUG, where the SPI is located) and `DVC' (sector 5), marked in red in figure~\ref{fig:axuv_toroidal}, and the others, marked in black, are additional vertical-type cameras.
This type of camera consists of three independent pinholes, one for each diode array.
The other, so-called horizontal camera, is located near the outer midplane, and such cameras are present in two toroidal sectors of the device.
Figure~\ref{fig:axuv_toroidal} shows the horizontal cameras with green that are used in the paper, `DHT' (sector 16) and `DHC' (sector 5).
In this type of camera, all three diode arrays share a common pinhole.
Both types of cameras have full line-of-sight coverage of the poloidal cross-section.
ASDEX Upgrade has 16 sectors in total, and the camera pair in sector 5 is toroidally $\sim$110 degrees away from the SPI.
The diodes in the cameras have a rectangular surface of 2\,mm\,$\times$\,5\,mm, with the longer side along the toroidal direction.
The aforementioned pinholes are rectangular slits with a poloidal width of 0.8 mm in both camera types, but the toroidal width of the slit is larger (3 mm) in the horizontal cameras than in the vertical cameras (2 mm).

\begin{figure}
  \centering
  \includegraphics[width=0.5\textwidth]{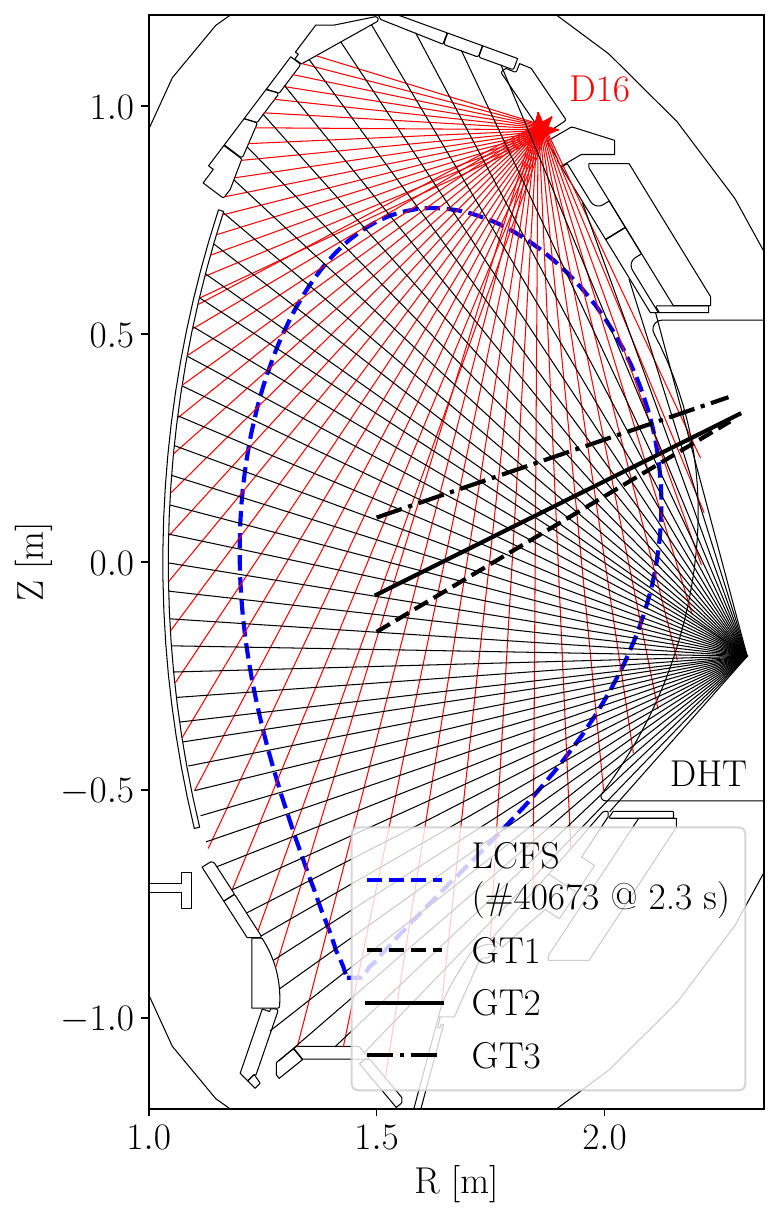}
  \caption{The poloidal cross-section of the ASDEX Upgrade tokamak with some of the AXUV lines of sight in sector 16, the separatrix (labelled LCFS ---last closed flux surface --- in the figure) of an SPI H-mode equilibrium (\#40673 @2.3 s), and the SPI injection directions set by the shatter heads for the specific guide-tubes (`GT').
  The diode numbers increase clockwise for both the vertical `D16' and horizontal `DHT' AXUV cameras.
  }\label{fig:2-axuv_los_poloidal}
\end{figure}

Figure~\ref{fig:2-axuv_los_poloidal} shows the lines of sight in the poloidal cross-section of AUG for the vertical camera with black and for the horizontal camera with dark red.
Additionally, the separatrix of a typical SPI H-mode equilibrium (AUG \#40673 @2.3 s) is also plotted with a dashed blue line.
The three straight lines pointing from the LFS towards the magnetic axis are the SPI vectors of the three different 2022 shatter head configurations in AUG~\cite{dibon_design_2023, Heinrich_2025_PhD}.
These show the shattering planes of the shatter heads, as seen in figure 5 of~\citep{dibon_design_2023}.

This coverage makes the cameras optimal for comparing SPI experiments with 3D SPI simulations.
In two toroidal sectors of the tokamak, both vertical and horizontal cameras are present, enabling some degree of cross-sectional comparisons of the radiation at two toroidal locations during shattered pellet injection.

\section{Single line of sight simulation}\label{sec:single-LOS}

\begin{figure}
  \centering
  \includegraphics[width=0.8\textwidth]{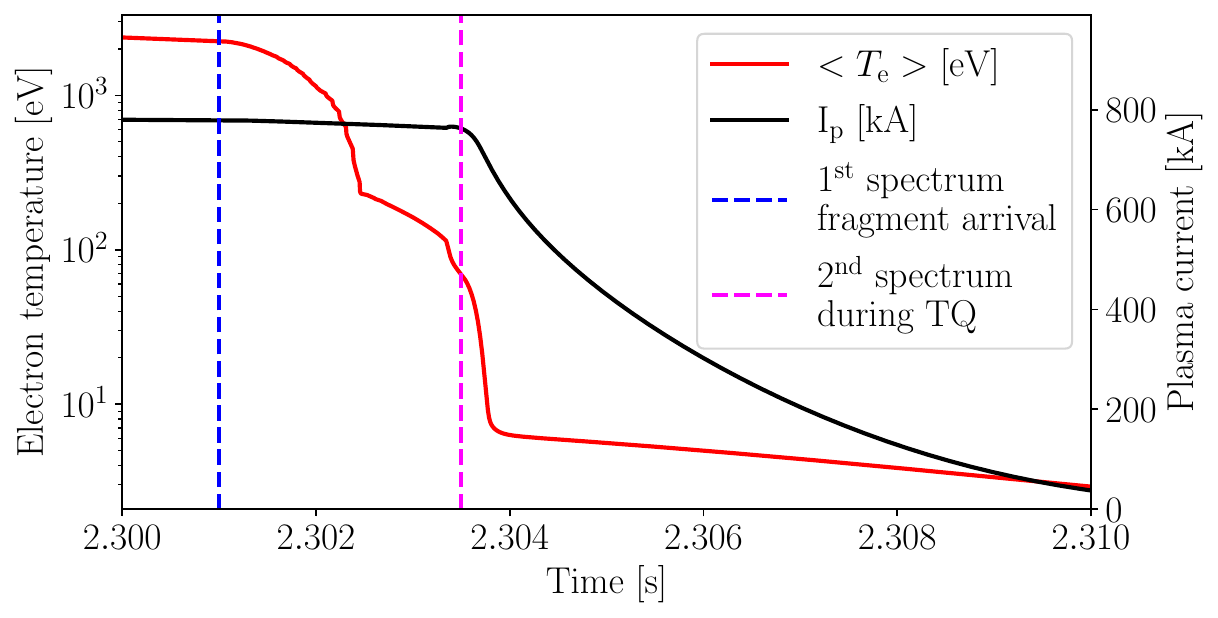}
  \caption{Plasma current and radially-averaged electron temperature time evolution from the spatially 1D DREAM SPI simulation containing a 10\,mol\% Ne / 90\,mol\% D$_2$ pellet corresponding to AUG discharge \#40949.
  The blue dashed line indicates the arrival of the fragments in the plasma, and the magenta line indicates the start of the thermal quench~\citep{halldestam_reduced_2025}.}\label{fig:3-DREAM_time_ev}
\end{figure}

\begin{figure}
  \centering
  \includegraphics[width=0.85\textwidth]{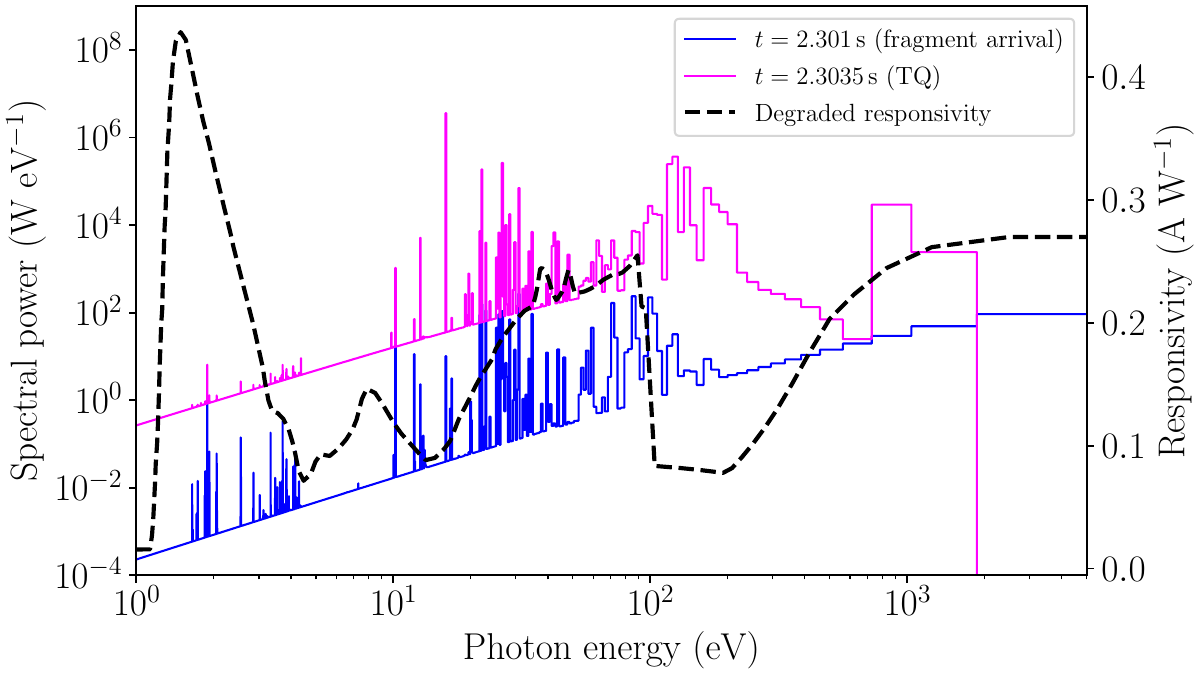}
  \caption{DREAM SPI emitted spectra taken at different times (see figure~\ref{fig:3-DREAM_time_ev}) during the simulation with blue and magenta.
  The overlaid dashed line shows the average degraded spectral response function.}\label{fig:4-DREAM_spectra}
\end{figure}

Before developing a 3D model, a study of the detection efficiency of AXUV diodes during mixed Ne/D$_2$ SPI was conducted to evaluate the effects of uncertainties in the spectral response function on the effective response of AXUV detectors in the case of dynamically evolving spectra~\citep{lengyel_plasma_2024}.
A single LOS was used as a representative observer to determine the detected spectral power density in a spatially 1D plasma simulation chosen from the series of DREAM~\citep{hoppe_dream_2021} simulations published by Halldestam et al~\citep{halldestam_reduced_2025}.
The LOS viewed the middle of a cylindrical plasma column modelled in Cherab using DREAM simulation outputs, and synthetic measurements were taken at different times during the simulation.
The chosen DREAM simulation from~\citep{halldestam_reduced_2025} is intended to reproduce SPI H-mode AUG discharge \#40949, during which one 10\,mol\% Ne / 90\,mol\% D$_2$ shattered pellet was injected (8~mm diameter pellet, $v_{\mathrm{inj}}=214.6\,\mathrm{ms^{-1}}$, $N_{\mathrm{fragments}}=72$, shattering angle $12.5^\circ$).
The time evolution of the plasma current and average electron temperature is shown in figure~\ref{fig:3-DREAM_time_ev}, as simulated in DREAM by Halldestam~\citep{halldestam_reduced_2025}.
The two vertical dashed lines denote chosen times for which the spectral power is plotted in figure~\ref{fig:4-DREAM_spectra}.

From the simulation output, Cherab was supplied with ion charge-state densities, as well as temperature and electron density profiles to compute the emitted radiation at different times during the examined time frame.
Two example spectra --- taken at different time steps during the simulation --- are shown in figure~\ref{fig:4-DREAM_spectra} overlaid with the degraded spectral responsivity function.
The included emission models were Bremsstrahlung and line radiation for neon and deuterium from OPEN-ADAS~\citep{summers_adas_2004}.
Note that no background impurities were included in these simulations~\citep{halldestam_reduced_2025}.
At the neon content considered here, the disruption radiation is nevertheless dominated by the injected neon.
Since Raysect is primarily focused on the visible spectrum, it uses equidistant spectral binning in wavelength.
Therefore, to achieve sufficiently high energy resolution in the XUV and SXR regions, 2500 spectral bins were used over the 1~eV~--~5~keV range (corresponding to $\sim$1240~nm and $\sim$0.25 ~nm, respectively), yielding a spectral resolution of $\sim$0.5 ~nm.

The Raysect \texttt{SightLine} object was used for observation with \texttt{SpectralPowerPipeline} to measure the spectral power $P(\lambda,V)$.
$P$ depends on the wavelength $\lambda$, and volume $V$.
The spectral power values obtained at different time steps throughout the discharge were applied as weights to the $R(\lambda)$ spectral responsivity functions shown in figure~\ref{fig:spectral_response}, to determine the $\eta_\text{eff} \left[A/W\right]$ spectrum-dependent effective response of the diodes.
We can index the spectral bins with $i$, and then the expression becomes:

\begin{equation}
    \eta_\text{eff}=\frac{\sum_i P_i(V)\cdot R_i}{\sum_i P_i(V)}.
\end{equation}\label{eq:eta_eff}

\begin{figure}
  \centering
  \includegraphics[width=0.75\textwidth]{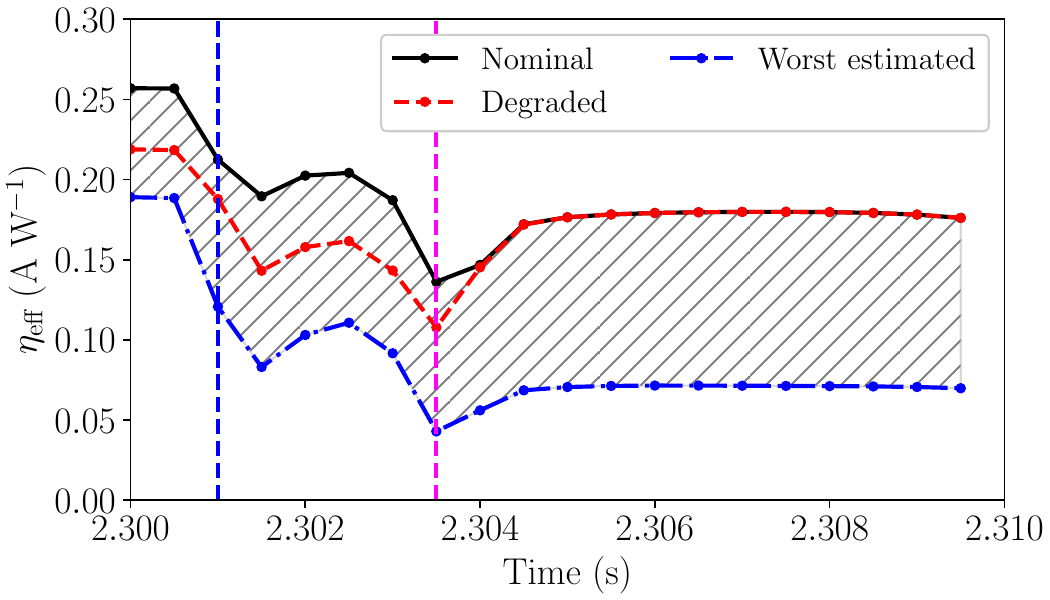}
  \caption{Time evolution of the effective response during the DREAM simulation.
  The solid black and red dashed curves correspond to nominal and degraded sensitivities, while the blue dash-dotted line to the worst estimation present in figure~\ref{fig:spectral_response} used to determine the effective response.
  The vertical dashed lines mark the times at which the radiation spectra are plotted in figure~\ref{fig:4-DREAM_spectra}.
  The grey hatched region shows an interval where the expected effective response of the AXUV diodes falls during mixed Ne/D$_2$ SPI-induced disruptions.}\label{fig:5-det_eff_te}
\end{figure}

Figure~\ref{fig:5-det_eff_te} presents the two fundamental challenges when trying to use AXUV diodes for determining absolute radiated power from mixed Ne/D$_2$ SPI experiments.
The first is the effect of degradation, which causes differences in spectral efficiency relative to the nominal values (red vs. black line in the figure) while the thermal quench has not yet occurred.
This difference is also highly uncertain.
This affects the measurements in any plasma discharge, not just in SPI experiments.
The second challenge arises from significant changes in the spectra of the emitted radiation during the time evolution of mixed Ne/D$_2$ SPI experiments.
These changes are most noticeable around 100 eV, where the bump in the spectrum shifts to higher energies as the plasma cools, and the spectral responsivity drops by more than a factor of 2.
Additionally, as the temperature decreases, Bremsstrahlung is significantly reduced in the high-energy range, and neon line radiation instead appears around 1 keV.
Since responsivity varies significantly with photon energy, this spectral change directly affects the calculated $\eta_{\text{eff}}$ effective response.

These considerations, together with a preliminary study~\citep{lengyel_plasma_2024}, led us to conclude that tomography in this context was not straightforward.

\section{3D synthetic diagnostic}\label{sec:3D-synth-diag}

To enable detailed comparisons between experimental data and plasma simulations, a 3D synthetic diagnostic was developed for different AXUV cameras in AUG~\citep{lengyel_plasma-sdsbolometeraxuv_2026}.
Cherab served as the basis for the synthetic diagnostic, and this section provides details on the technical implementation.

The diodes are placed in 3D space as Cherab \texttt{BolometerFoil} objects, inside \texttt{CameraBox} objects with rectangular slits (\texttt{BolometerSlit}) cut out.
The diodes have a sensitive area of 2\,mm\,$\times$\,5\,mm~\citep{bernert_application_2014}.
In cameras that view the poloidal cross-section (all cameras covered in this article), the longer sides are aligned with the toroidal direction.
Geometry data is provided in AUG shotfiles, and users of the synthetic diagnostic need access to it to run simulations for AUG.
The toroidal slit dimension is 3 mm for cameras DHT and DHC, and 2 mm for D16 and DVC.
The poloidal dimension of the slits is calculated from the slit area in the AUG shot files and is approximately 0.8 mm for all cameras mentioned.

\begin{figure}
    \centering
    \includegraphics[width=\linewidth]{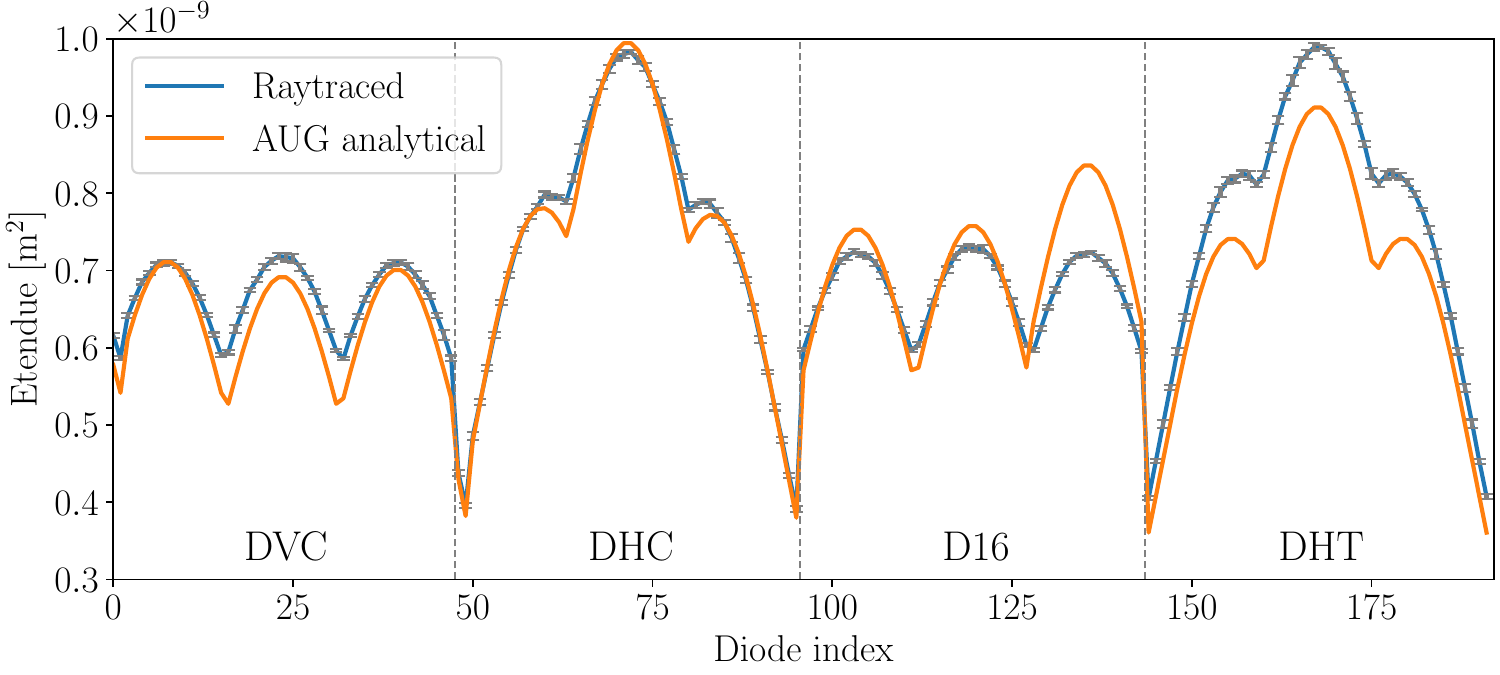}
    \caption{Comparison of the Cherab ray-traced etendue values of the synthetic diagnostic to the AUG analytically calculated values.
    Camera names are as introduced in section~\ref{sec:AUG-AXUV}.}\label{fig:6-etendue-comparison}
\end{figure}

At AUG, the $F$ etendue values are analytically calculated for the diodes based on the measured areas of the camera slits ($f_{\text{slit}}$) and the diode areas ($f_{\text{diode}}$).
The line-integrated power density $I$ is calculated from the power $P$ absorbed by the diode as
\begin{equation}
    I = P/F \quad \text{with} \quad F = \cos(\alpha)\, f_{\text{diode}}\, \Omega / (4\pi)
    \quad \text{and} \quad \Omega = f_{\text{slit}} \cos(\gamma) / d^2,
\end{equation}
where $d$ is the distance of the diode from the slit, $\alpha$ is the angle between the optical axes of the slit and the diode, and $\gamma$ is the angle enclosed by the diode normal vector and the line between the slit centre and the diode centre.
The comparison of the analytic and ray-traced etendue values is shown in figure~\ref{fig:6-etendue-comparison}.
The mean relative difference is 2.3\%, and the maximum relative difference is 7.09\%.
The differences mainly stem from the fact that the ray-tracing approach can account for half-shadows, whereas the analytical formula can not~\citep{carr_description_2018}.

The volume from which radiation is expected is defined using a Cherab \texttt{RayTransferCylinder} object, which can be regarded as an axisymmetric toroidal voxel grid with a user-defined resolution.
Using the \texttt{RayTransferPipeline}, later referred to as the ray transfer method ~\citep{kajita_usage_2016}, is recommended by the Cherab developers for observing multiple plasma states with the same observation geometry.
For each diode, $10^6$ rays are traced through its rectangular slit and through the voxel grid to compute the spatially resolved sensitivity matrices.
The simulation may include reflections from the realistic CAD geometry of the plasma-facing components (PFCs) if they are defined with the appropriate materials in Cherab.

Different emission data defined on the same voxel grid can then be multiplied by the diode sensitivity matrices to determine radiation power on the diode surfaces.
The radial and vertical sides of the voxels should be the same length, and the grids for the plasma emissions and the diode sensitivity simulation should be identical to avoid sampling issues.
In the cases presented here, we used a 2~cm $\times$ 2~cm resolution for both the plasma emission and sensitivity simulations in the radial and vertical dimensions.
A higher resolution of 1~cm $\times$ 1~cm was also tested without reflections. The difference from the lower resolution case was 5\% on average ($\sim$ 10\% maximum).
Note that the edge channels observing only a few cells of the voxel grid are affected the most by the resolution, and two channels on both edges of each camera were therefore excluded from the calculation of the statistics of the deviation.
During the SPI-induced disruptions, the AXUV signals vary by 4 orders of magnitude, thereby rendering a 5-10\% difference negligible.
Therefore, we concluded that it is safe to use the lower (2~cm $\times$ 2~cm) resolution version for our purposes.
The differences for the core channels mostly originate from the emission data when the input emission is inhomogeneous on length scales comparable to the voxel size.

\begin{figure}
    \centering
    \includegraphics[width=0.6\linewidth]{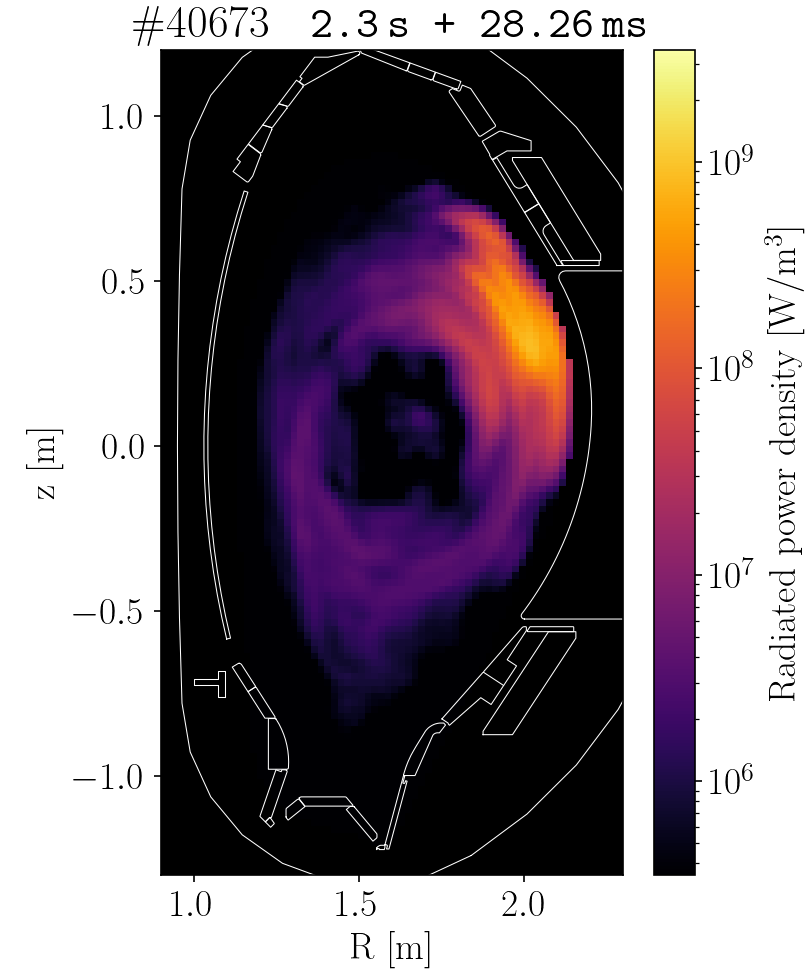}
    \caption{Example of emitted power density distribution from one of the analysed simulations. The figure is a frame from Movie 1, provided as supplementary data with the article, showing the time evolution of the radiation in the poloidal cross-section.}
    \label{fig:example_emission}
\end{figure}

The synthetic diagnostic is applied here to input data from JOREK~\citep{tang_non-linear_2025,tang_quantitative_2026,hoelzl_jorek_2021} defined on a custom 2D poloidal grid; therefore, the electron density, electron temperature, and impurity ion densities with different charge-states are super-sampled from the poloidal JOREK grid onto a 4$\times$ finer uniform grid in both radial and vertical dimensions.
Then, a Gaussian anti-aliasing filter is applied to the input data, and the resulting signal is sampled at the midpoints of the coarser grid to match the grid dimensions used for the diode sensitivity simulation.
An example of radiated power density distribution is shown in figure~\ref{fig:example_emission} from a simulation of AUG discharge \#40673, described in detail in subsection~\ref{subsec:synth-exp-comp}.

Two fundamentally different vessel geometries are implemented in the synthetic diagnostic.
The first geometry starts with a simple rectangle in the poloidal cross-section that encompasses the full extent of the plasma.
This rectangle is then toroidally rotated to form a hollow cylindrical container for the entire plasma volume, and its material is set to be fully absorbing.
In this case, when the traced rays reach this surface, the ray-tracing stops.
The second type of geometry is the full, real geometry of the AUG PFCs, which can be loaded from CAD files of the vessel.
Using the CAD geometry is essential when determining diode sensitivity with reflections, and the material of the PFCs can be chosen.
For this study, since AUG has mostly tungsten PFCs, all components are chosen to be tungsten with a roughness of 0.29, where roughness 0 is completely specular and roughness 1 results in fully diffuse reflections.
Note that during plasma discharges, surface conditions may change significantly, so this roughness value is used only as an estimate.

\begin{figure}
    \centering
    \includegraphics[width=0.6\linewidth]{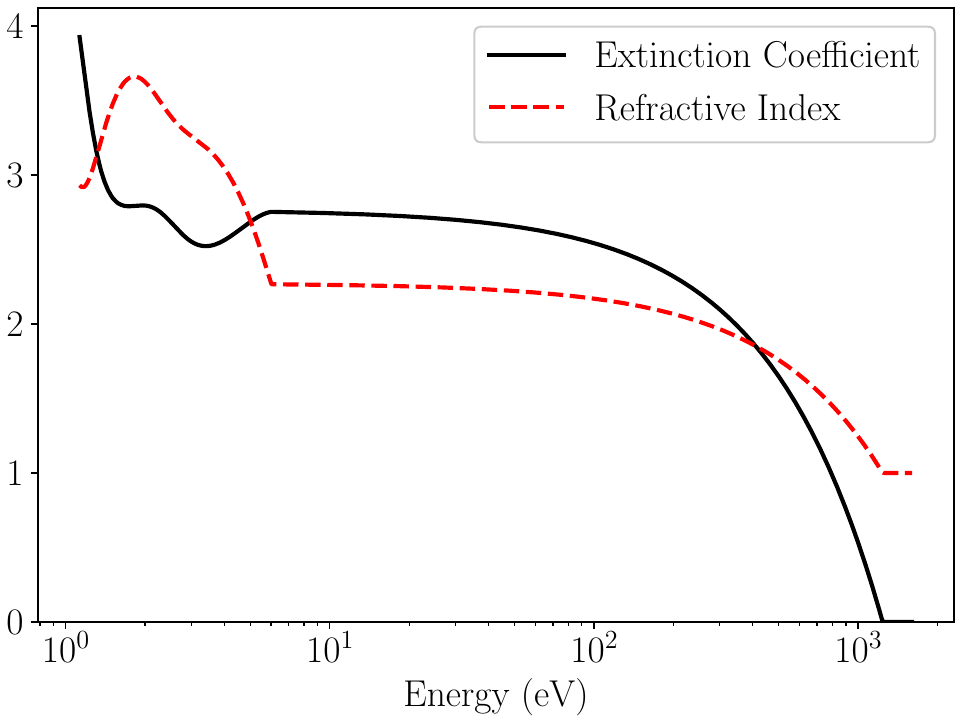}
    \caption{Tungsten refractivity and extinction coefficients used. Based on Rakic~\citep{rakic_optical_1998} (as used by Raysect) and extrapolated to $\left. n \right|_{1.24\text{ keV}}=1$ and $\left. k \right|_{1.24\text{ keV}}=0$.}
    \label{fig:tungsten_refractivity}
\end{figure}

\begin{figure}
  \centering
  \begin{subfigure}[b]{0.49\textwidth}
      \includegraphics[width=\textwidth]{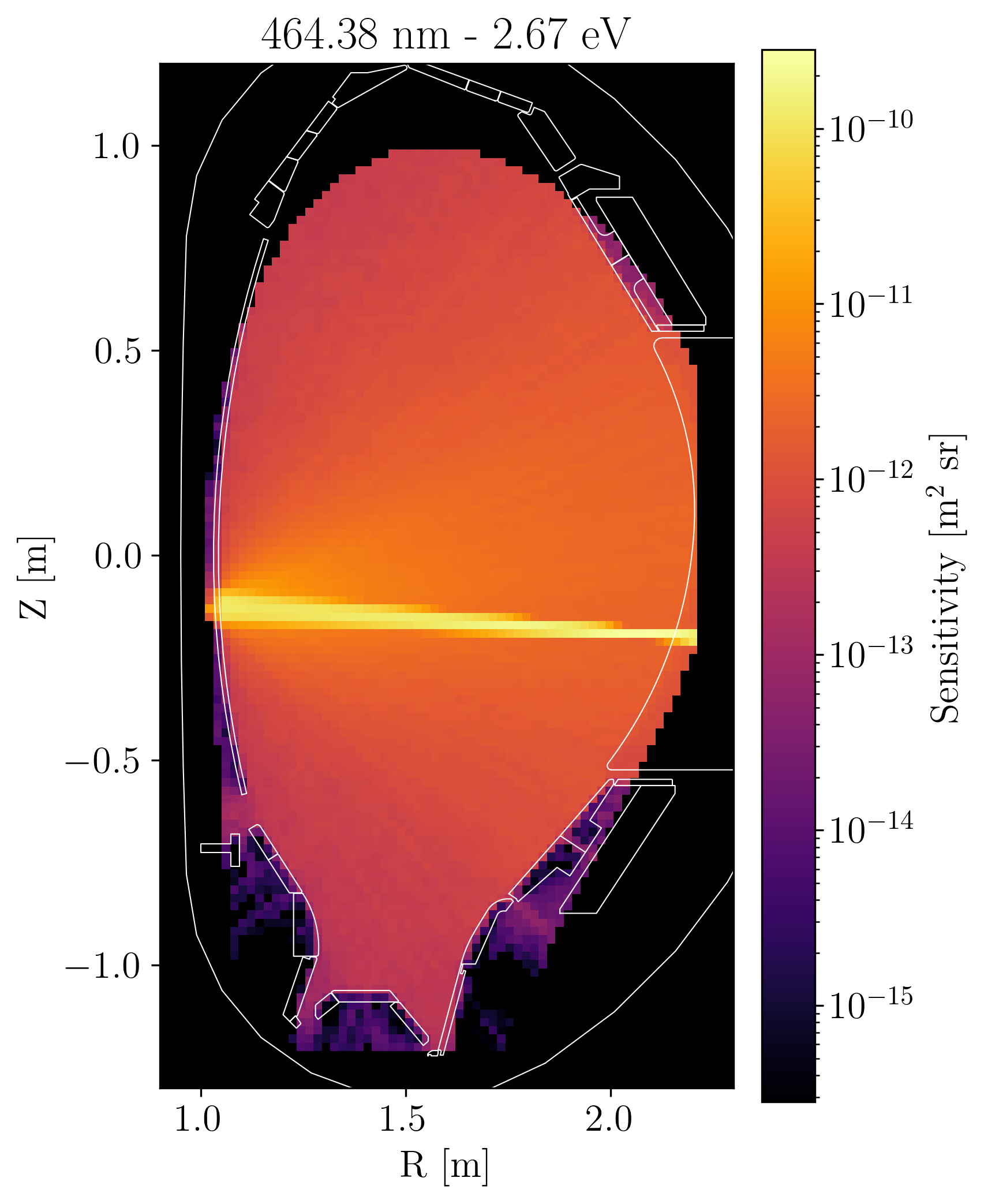}
      \caption{Visible}

  \end{subfigure}
  \begin{subfigure}[b]{0.49\textwidth}
      \includegraphics[width=\textwidth]{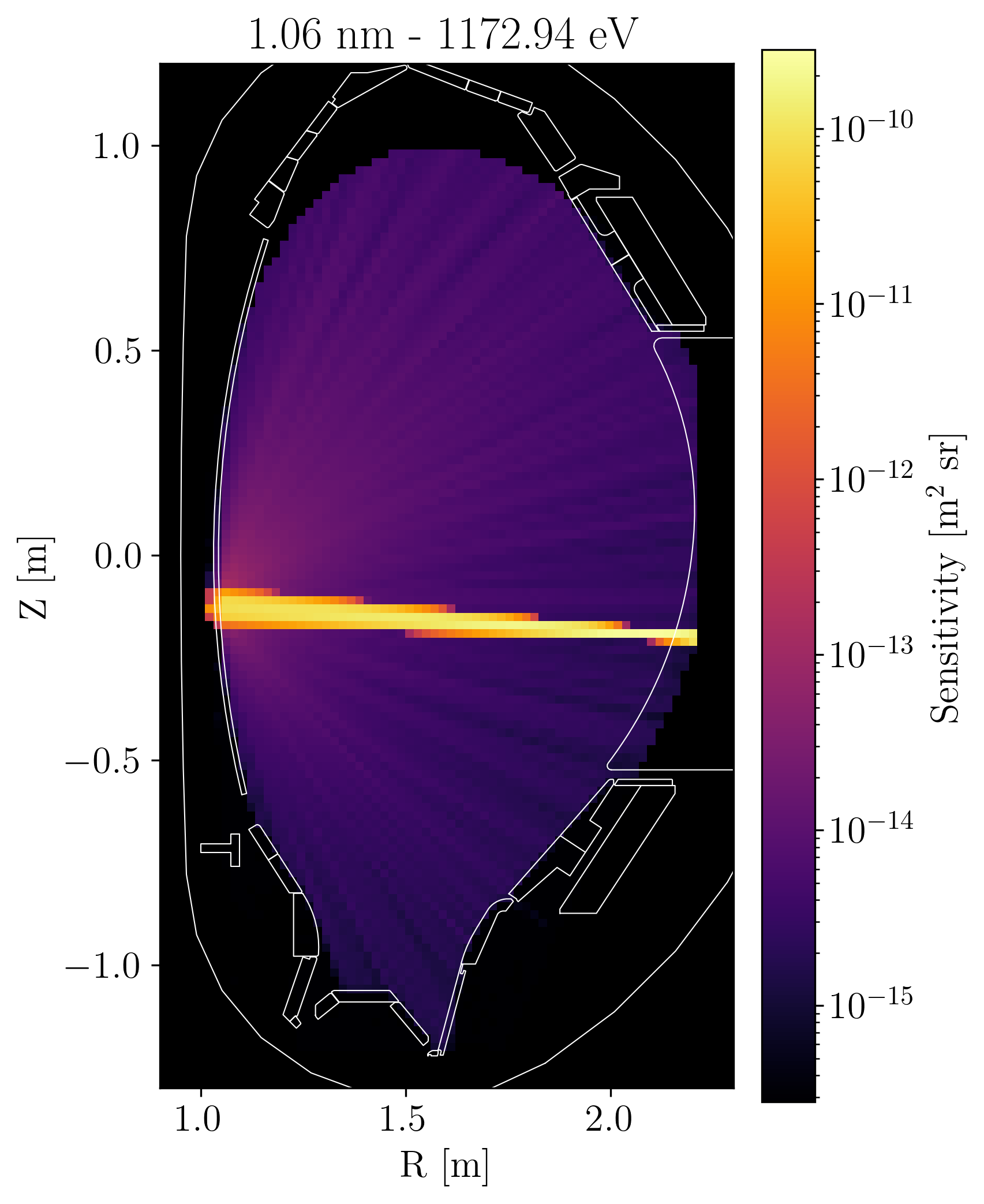}
      \caption{SXR}

  \end{subfigure}
  \caption{Visualisation of the sensitivity of chosen lines of sight with reflections from the plasma-facing components accounted for.
  Sub-figure (a) shows the values with which the specific toroidal voxels contribute to the sensitivity of a chosen LOS in the horizontal AXUV camera in sector 5 in the visible wavelength range.
  Sub-figure (b) shows the same contributions, but in the SXR range.
  }\label{fig:7-reflections}
\end{figure}

The reflections are calculated using wavelength-dependent parameters for the real and imaginary parts of the complex refractive index, denoted by the refractive index ($n$) and extinction coefficient ($k$), respectively.
For the visible range, the Raysect default values are used.
However, the examined spectrum extends to $1.24$ keV, and the coefficients towards the high-energy range are determined by linearly extrapolating from the last data point, assuming $\left. n \right|_{1.24\text{ keV}}=1$ and $\left. k \right|_{1.24\text{ keV}}=0$.
The assumption is based on the highest energy datapoint from Windt\citep{windt_optical_1988}, which was recommended by the Refractiveindex.info database of optical constants~\citep{polyanskiy_refractiveindexinfo_2024}.
The used refractive index and extinction coefficients are shown in figure~\ref{fig:tungsten_refractivity}.
Some examples of the resulting sensitivity matrices for chosen diodes and wavelengths are shown in figure~\ref{fig:7-reflections}.

\begin{figure}
    \centering
    \begin{subfigure}[b]{0.49\textwidth}
        \centering
        \caption{With reflections}
        \includegraphics[width=\textwidth]{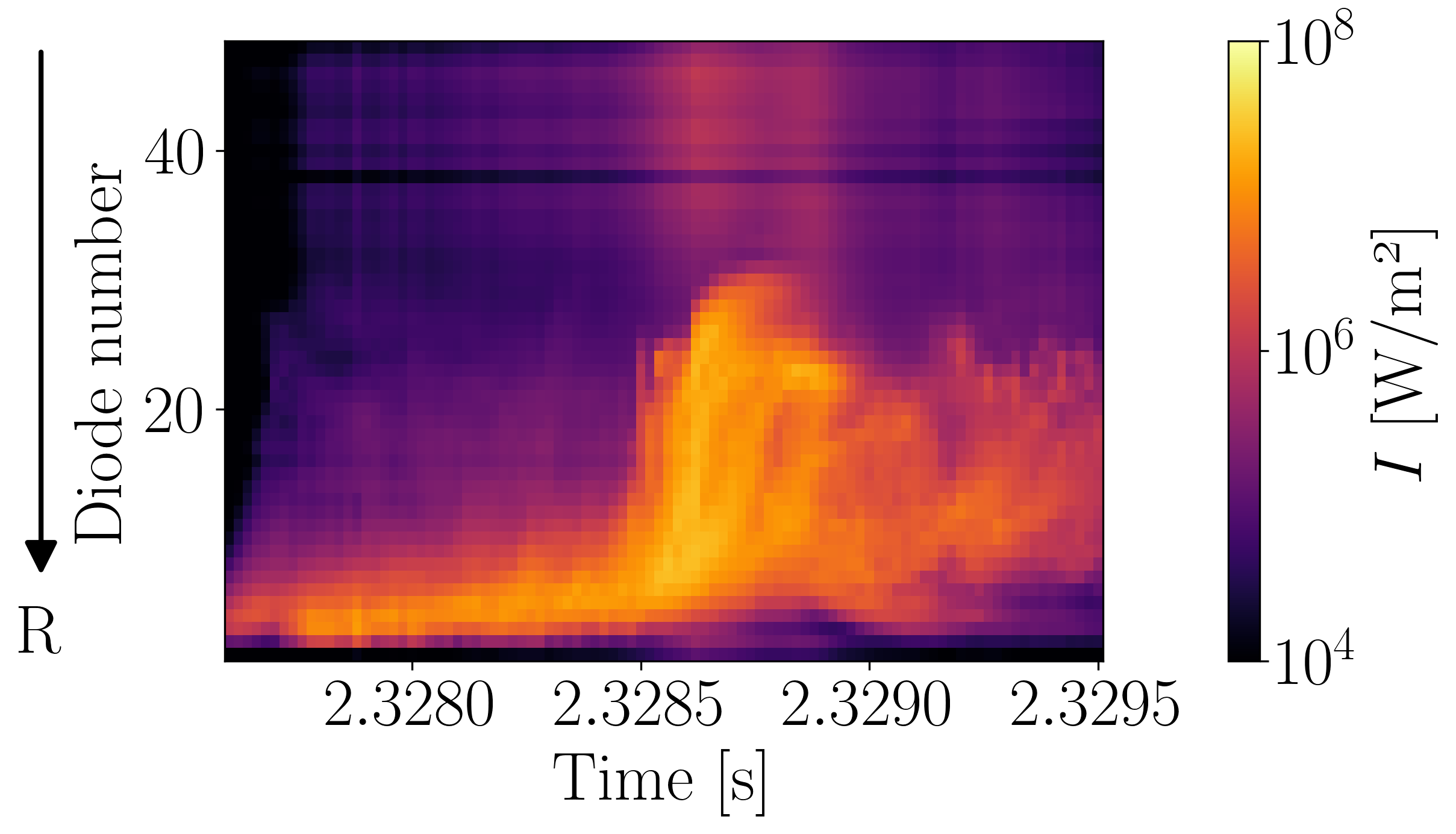}
        \label{fig:reflections-on}
    \end{subfigure}
    \hfill
    \begin{subfigure}[b]{0.49\textwidth}
        \centering
        \caption{Without reflections}
        \includegraphics[width=\textwidth]{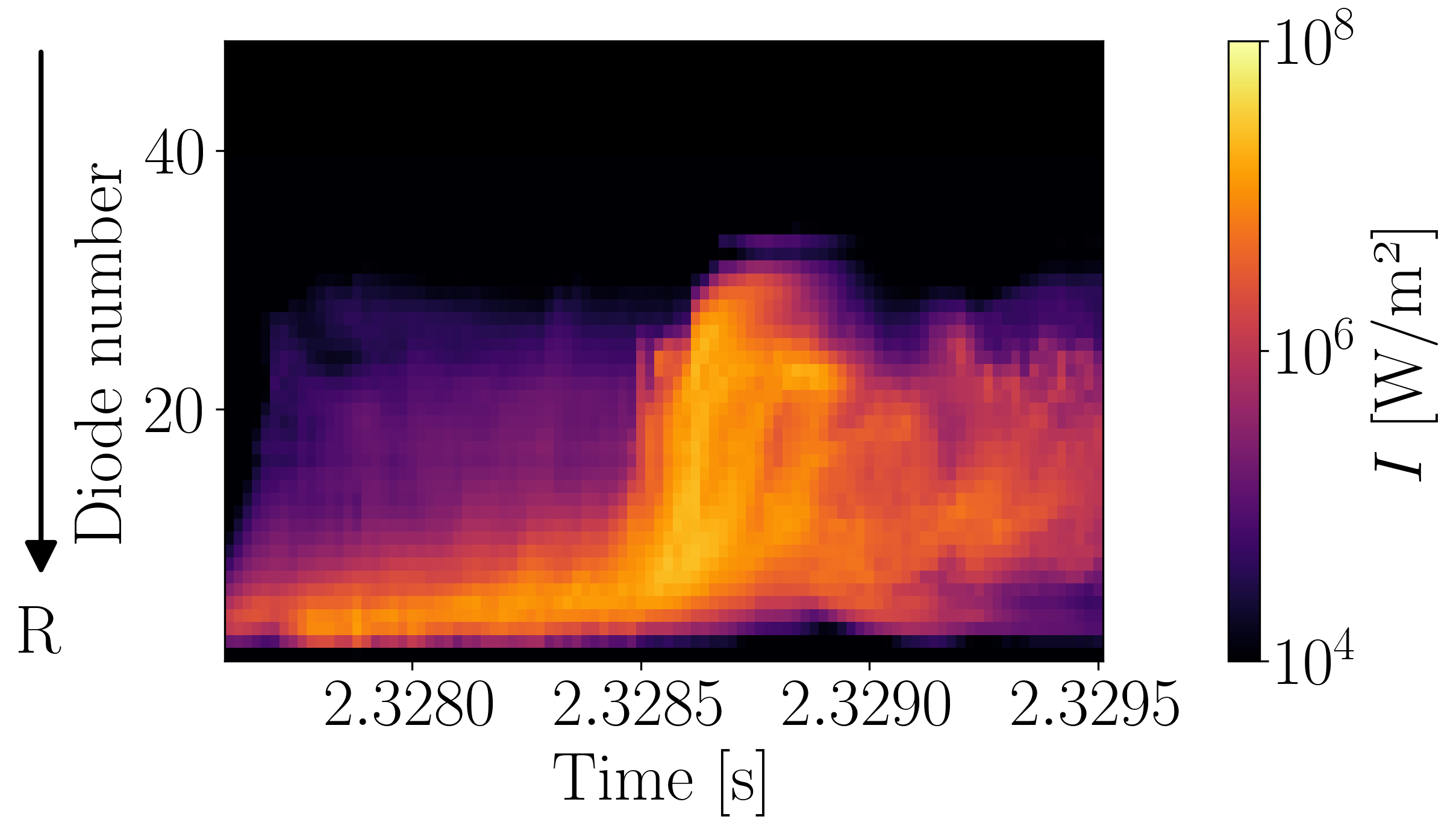}
        \label{fig:reflections-off}
    \end{subfigure}
    \caption{Illustration of the effects of reflections on the line-integrated brightness measured by the synthetic diagnostic for the sector 16 horizontal camera `DHT' during a JOREK SPI simulation.
    The figure on the left shows a synthetic measurement with reflections-enabled sensitivity matrices, and the figure on the right shows a synthetic measurement with no reflections used in the calculation of the sensitivity matrices.
    The reflection contribution is most apparent in the edge channels, but gives a general background for all diodes.}\label{fig:reflection-on-off-comparison}
\end{figure}

As mentioned earlier, surface conditions, including reflectivity, may change during discharges, so the reflection contribution can carry significant error due to poor knowledge of the surface properties.
A comparison of the line-integrated brightness signals for an example simulation is shown in figure~\ref{fig:reflection-on-off-comparison}.
It is worth noting that the LOS of diode number 38 observes a gap between PFCs on the inboard side (see figure~\ref{fig:7-reflections}), resulting in significantly less reflected radiation being detected by that diode.
Additionally, some diode lines of sight near the edges do not pass through the plasma volume.

The upper spectral boundary was lowered from 5~keV in the single-LOS simulation to $\sim$1.24~keV (1~nm), since this paper focuses on simulations and measurements of SPI-induced disruptions.
In disruptions, the $>1$~keV range is only relevant at the very beginning, while the temperature is still above this value, and Bremsstrahlung is emitted in that spectral range.
Additionally, neon does not radiate in that high-energy range, as shown in figure~\ref{fig:4-DREAM_spectra}.
To further reduce the computational cost of ray tracing, rather than linearly binning the entire examined spectrum in wavelength, the full spectrum was divided into 3 regions: 1--10~eV, 10--100~eV, and 100--1240~eV.
All 3 regions were then divided into 100 bins, yielding 300 total spectral bins for the 3D synthetic diagnostic, compared with the 2500 spectral bins used in the previous, single-LOS simulation.
Finally, plasma emissions were supplied to the synthetic diagnostic using Cherab to calculate emissions from the post-processed JOREK simulation output, using the same spectral bins as in the diode sensitivity simulation.
For comparison with experiments, recent 3D SPI simulations~\citep{tang_quantitative_2026} were chosen, which did not include neutral deuterium; therefore, only Bremsstrahlung and neon line radiation were included in the emission simulations.

\subsection{Verification of the synthetic diagnostic}\label{subsec:verif}
Verification of the LOS geometry has been performed against the AUG data; the diode lines of sight overlap with those exported from the shotfiles.
The etendues of individual diodes have been verified against the analytically calculated values used at AUG, and a good match was found, with a mean relative difference of 2.3\%, as presented in figure~\ref{fig:6-etendue-comparison}.

Verification of the ray transfer method for the synthetic diagnostic has been performed against another method that Cherab can use to simulate the power reaching the diode surfaces.
Using the \texttt{PowerPipeline} method, instead of calculating emissions within the voxels, the plasma volume is sampled along rays traced from the diodes.
Homogeneous radiation was set up using the same voxel geometry as in the ray transfer method, and the average relative error in the observed power of the diode channels was 0.017\%, with a maximum relative error of 2.44\%.

Several factors hinder validation against standard plasma experiments.
The tungsten impurity concentration is difficult to measure accurately~\citep{angioni_impact_2025}, and tungsten radiates significantly in the VUV-SXR range~\citep{putterich_calculation_2010}, where the sensitivity of the diodes degrades significantly during plasma operation~\citep{bernert_application_2014}.
Additionally, tungsten atomic data is incomplete, and still subject of active research~\citep{peyrusse_tungsten_2026} due to the large number of its electrons, and these uncertainties cascade into the AXUV synthetic diagnostic.

Therefore, we opted to examine mixed Ne/D$_2$ SPI simulations and experiments in which the intrinsic impurity radiation is significantly lower than that of the injected impurity.
These SPI discharges have the advantage that the radiation structures are very clearly visible in the signals when plotted over time.

\subsection{Comparison of synthetic signals with experimental measurements}\label{subsec:synth-exp-comp}

Non-linear 3D JOREK~\citep{hoelzl_jorek_2021} SPI simulations were performed for two chosen AUG discharges with results that show a good quantitative match to several plasma parameters and their time evolution~\citep{tang_quantitative_2026}.
These simulations model the magnetohydrodynamic evolution of the plasma, including pellet-fragment ablation dynamics and impurity atomic processes (ionisation, recombination, and radiation), calculated using Monte Carlo markers.
The JOREK SPI simulation results from Tang et al~\citep{tang_quantitative_2026} are examined in this section using the synthetic diagnostic and compared with the corresponding discharges from the 2022 AUG SPI campaign.
For the comparisons, the degraded spectral sensitivity function was used (see figure~\ref{fig:spectral_response}), and reflections were switched on.

The AUG discharge \#40673 was an SPI experiment (equilibrium separatrix shown in figure~\ref{fig:2-axuv_los_poloidal}), in which the pellet contained 10\,mol\% Ne and 90\,mol\% D$_2$ (8~mm diameter pellet, $v_{\mathrm{inj}}=221.3\,\mathrm{ms^{-1}}$, shattering angle $25^\circ$).
Observations from experiments and modelling suggest that at higher neon contents relevant for this discharge, both the plasmoid (pellet cloud) drift and the rocket effect affecting the pellet fragments are significantly lower than for very low neon content pellets~\citep{Pegourie_2007,kong_2026_plasmoid_drift,vallhagen_drift_2023,corbett_numerical_2026,matsuyama_neutral_2022}.
Note that the rocket effect also heavily depends on the fragment size and velocity distribution.
However, due to the high neon content, we can still expect the fragment trajectories to approximately match the trajectory of the radiating pellet cloud.

\begin{figure}
  \centering
  \includegraphics[width=0.75\textwidth]{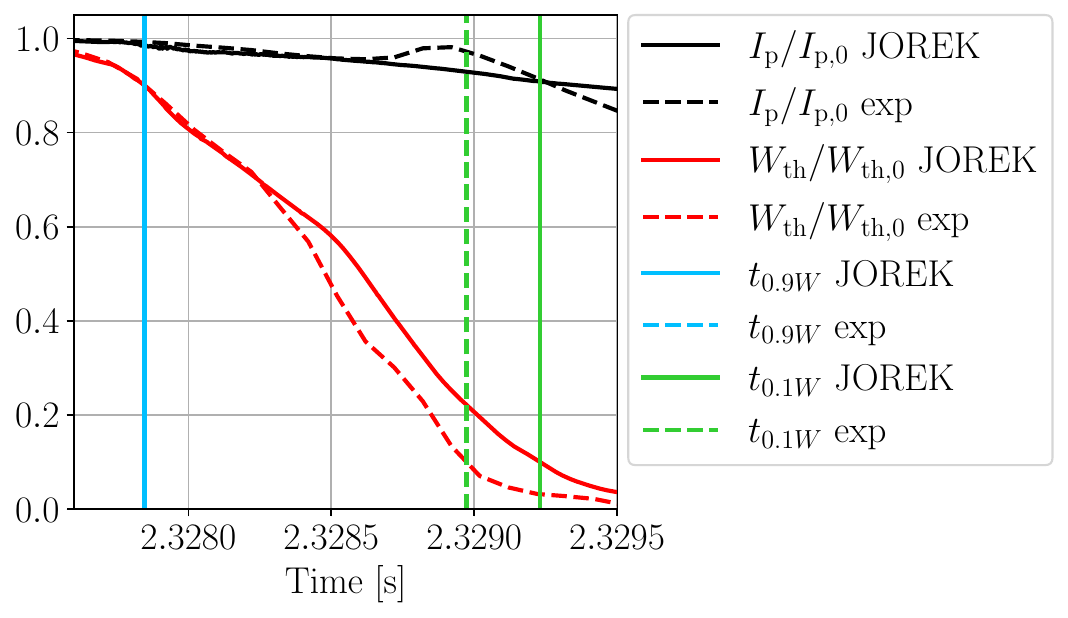}
  \caption{The time evolution of normalised plasma parameters from the AUG experiment \#40673 (10\% Ne), and of the corresponding JOREK simulation (based on figure 9 of \citep{tang_quantitative_2026}).
  Solid lines represent the simulation data, and dashed lines the experimental data.
  The black lines denote the normalised plasma current $I_{\mathrm{p}}$, and the red lines the normalised thermal energy $W_{\mathrm{th}}$.
  The 90\% and 10\% thermal energy points, $t_{0.9W}$ and $t_{0.1W}$, are marked with dashed lines for the experiment and solid lines for the simulation data.
  }\label{fig:40673_W_th}
\end{figure}

The time evolution of some of the normalised experimental data and that of the corresponding JOREK simulation is presented in figure~\ref{fig:40673_W_th}.
All the plotted parameters show a very good match between the JOREK simulation and the experiment.
It is also informative to examine the times at which the thermal energy is at 90\% and 10\%, $t_{0.9W}$ and $t_{0.1W}$ respectively, which we use as markers for the start and (almost) end of the significant thermal energy loss, providing a basis for a physics-based comparison of the experiments and the simulations.
The evolution of thermal energy $W_{\mathrm{th}}$ is strongly influenced by radiation-driven loss, as measured by the AXUV diagnostic, making these reference times useful for our comparisons between experiments and simulations.

\begin{figure}
    \centering
    \begin{subfigure}[b]{0.49\textwidth}
        \centering
        \caption{Synthetic, sector 16, vertical `D16'}\label{fig:40673-S16:a}
        \includegraphics[width=\textwidth]{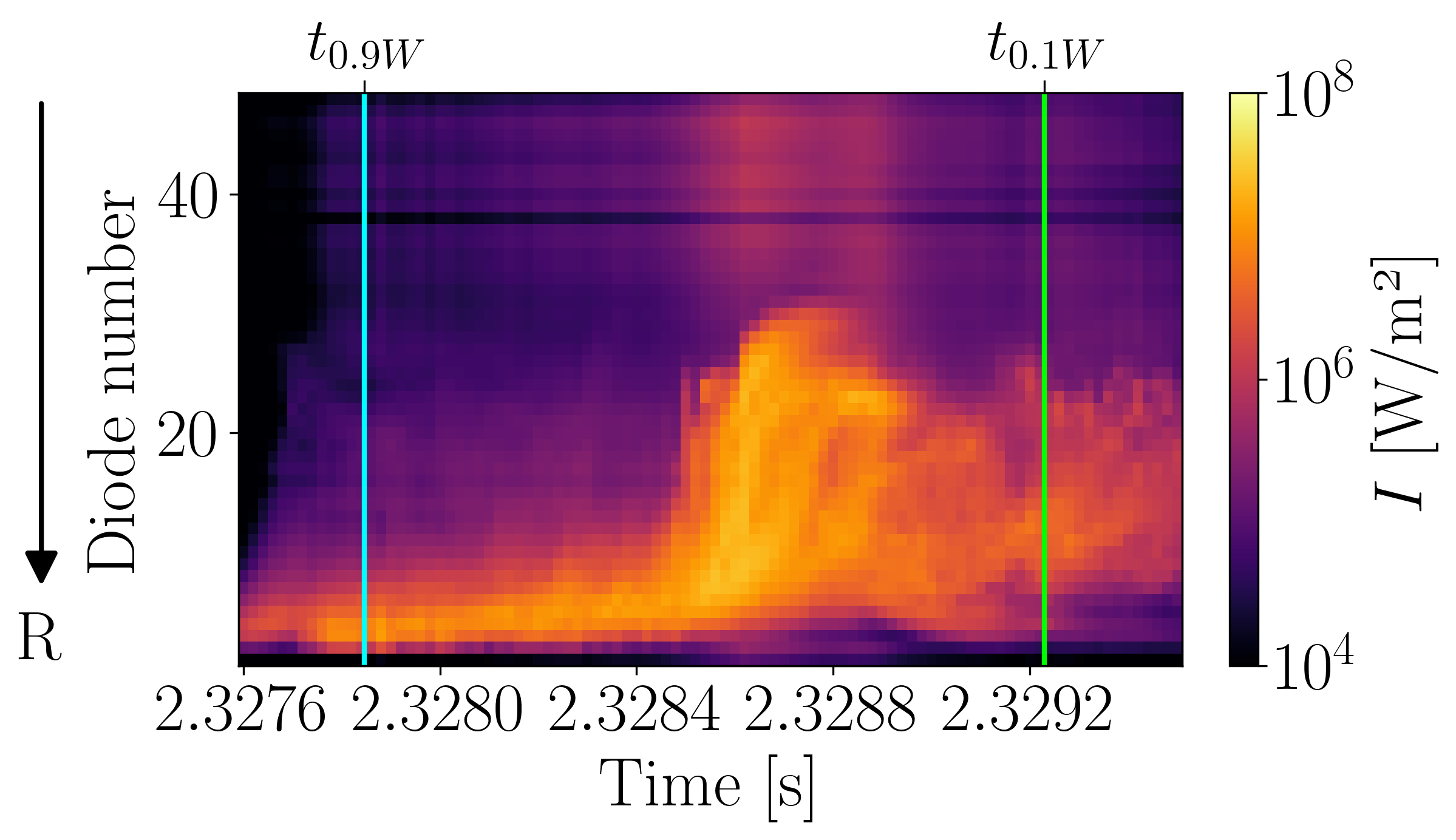}
    \end{subfigure}
    \hfill
    \begin{subfigure}[b]{0.49\textwidth}
        \centering
        \caption{Synthetic, sector 16, horizontal `DHT'}\label{fig:40673-S16:b}
        \includegraphics[width=\textwidth]{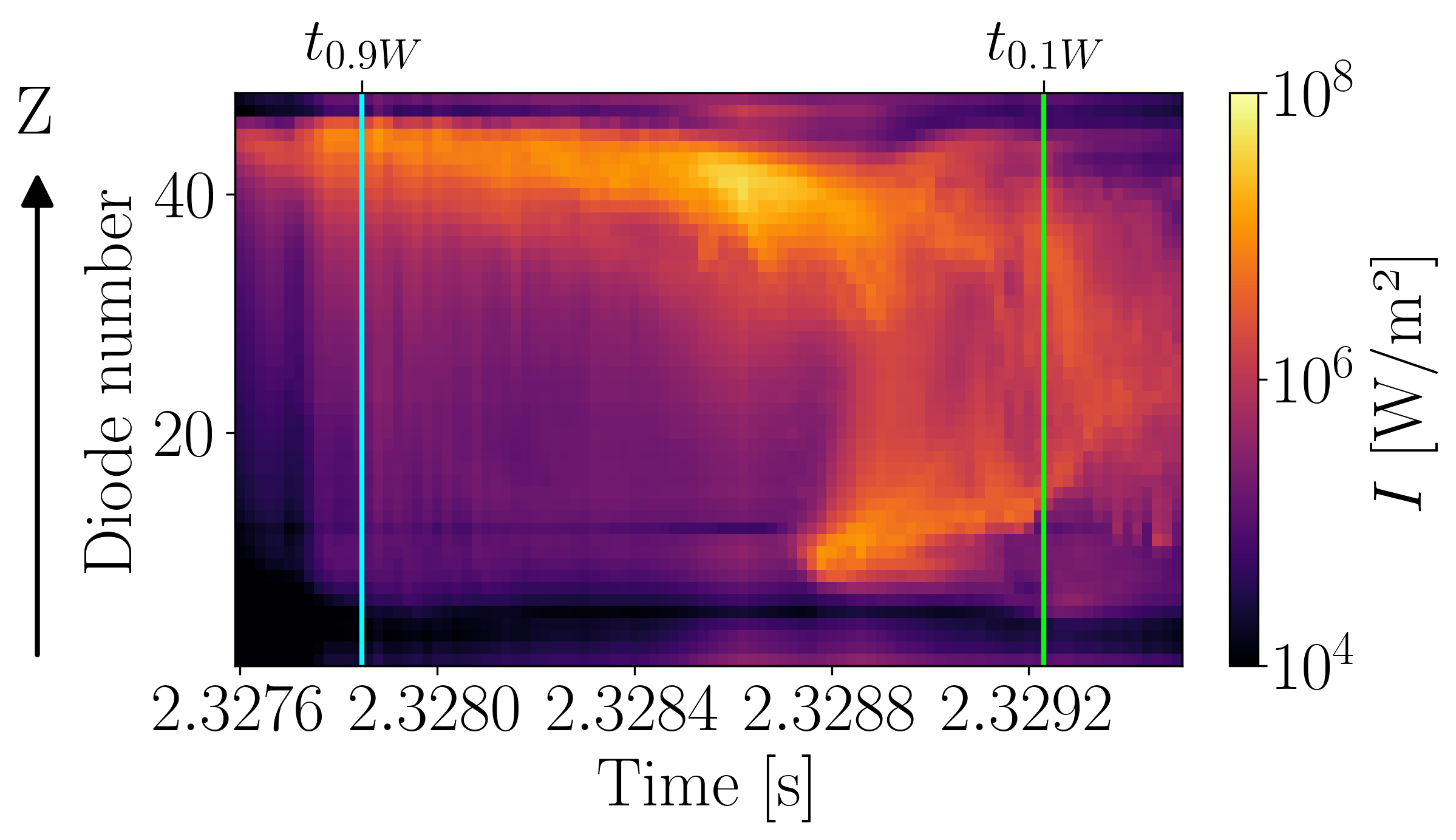}
    \end{subfigure}

    \begin{subfigure}[b]{0.49\textwidth}
        \centering
        \caption{Experimental, sector 16, vertical `D16'}\label{fig:40673-S16:c}
        \includegraphics[width=\textwidth]{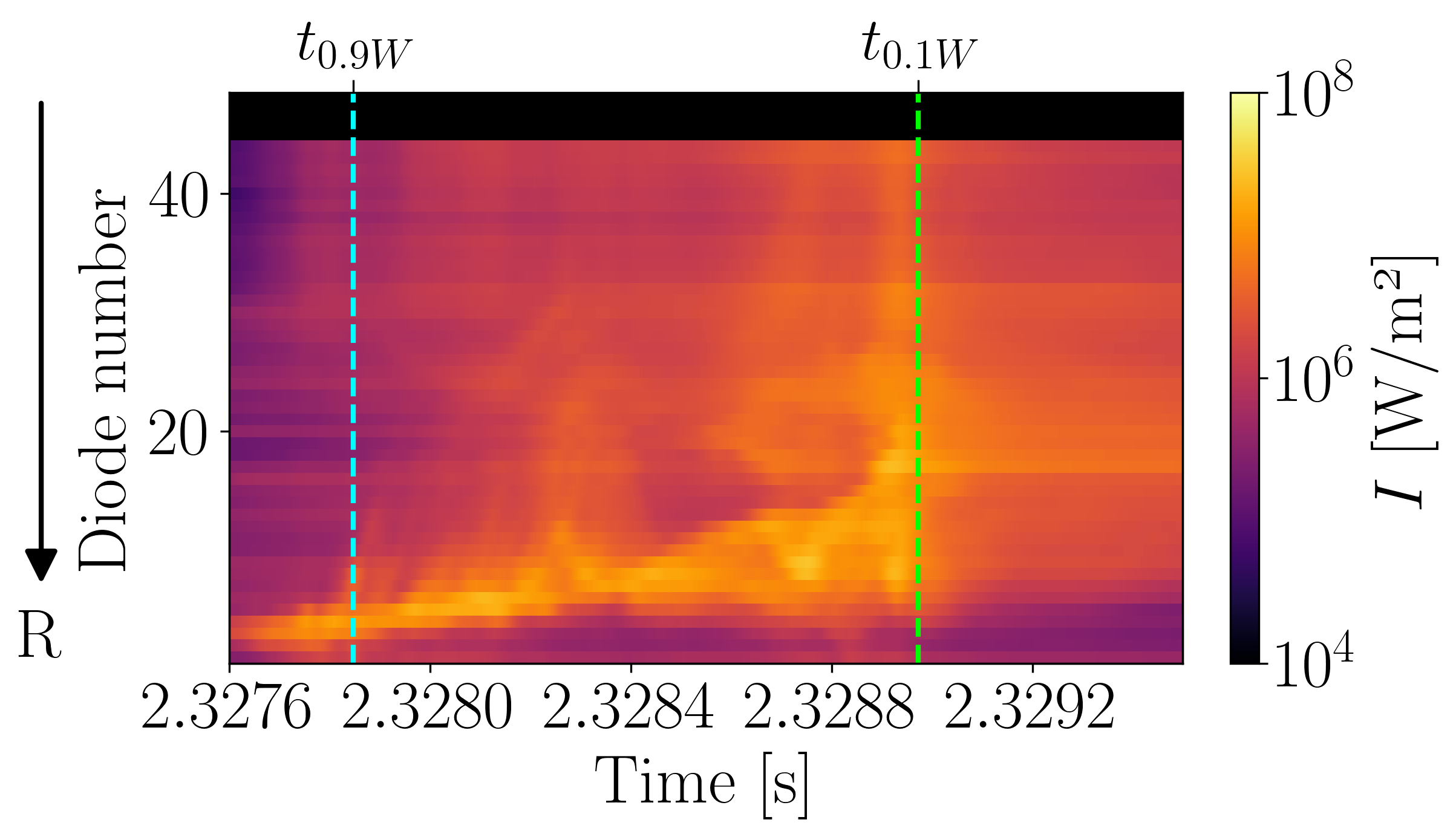}
    \end{subfigure}
    \hfill
    \begin{subfigure}[b]{0.49\textwidth}
        \centering
        \caption{Experimental, sector 16, horizontal `DHT'}\label{fig:40673-S16:d}
        \includegraphics[width=\textwidth]{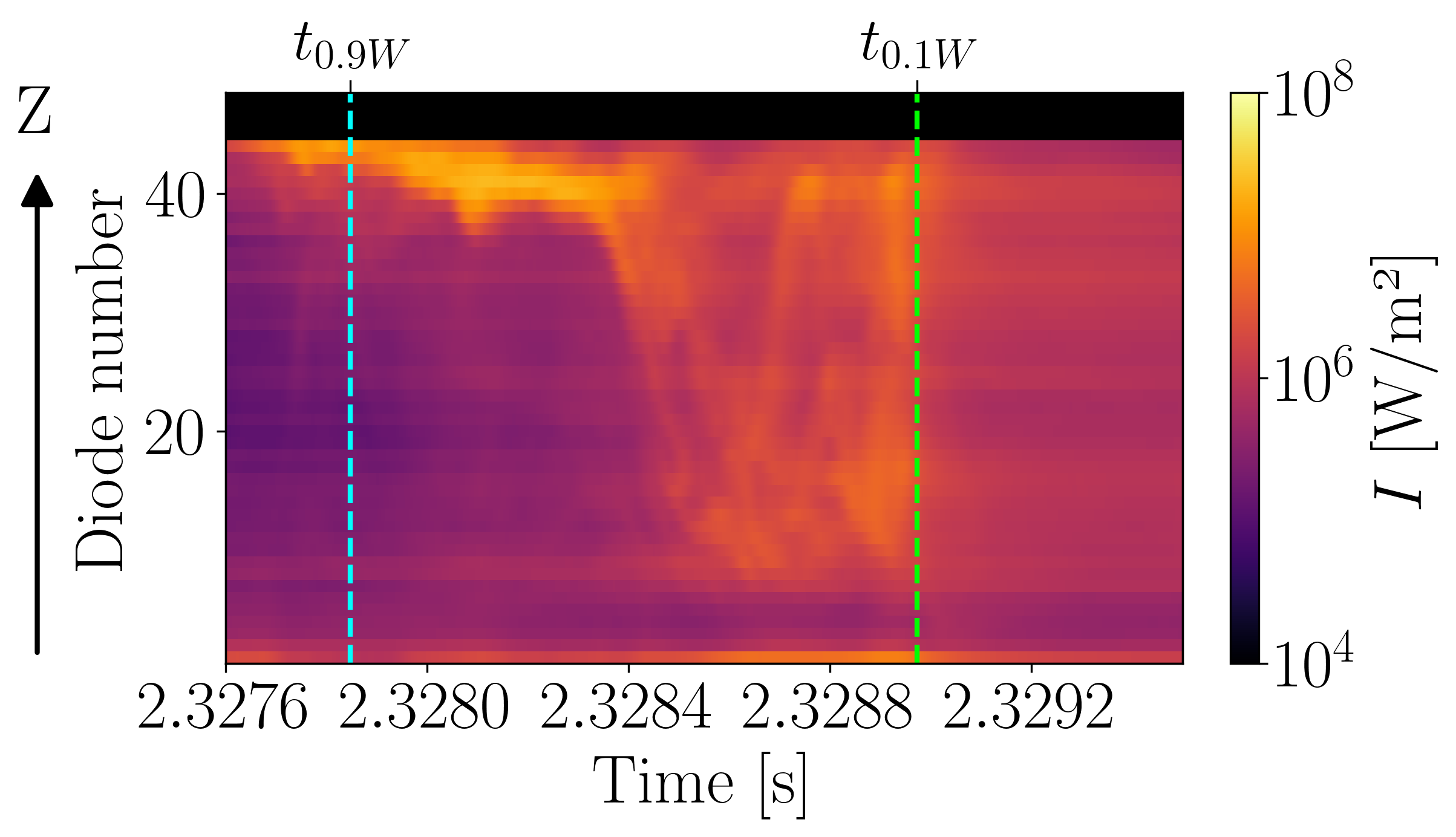}
    \end{subfigure}

    \caption{Comparison of synthetic and experimental AXUV data in sector 16 (sector of the SPI) for AUG discharge \#40673.
    The shattered pellet contained 10\,mol\% Ne.
    Figures (a) and (b) show the results of the synthetic measurement performed on the JOREK data for the vertical and the horizontal cameras `D16' and `DHT'.
    Figures (c) and (d) present the experimental data for the corresponding cameras (see figure~\ref{fig:2-axuv_los_poloidal}).
    The $t_{0.9W}$ and $t_{0.1W}$ times are marked with solid lines for simulation data, and with dotted lines for experimental data.
    }\label{fig:40673-S16}
\end{figure}

\begin{figure}
    \centering
    \begin{subfigure}[b]{0.49\textwidth}
        \centering
        \caption{Synthetic, sector 5, vertical  `DVC'}\label{fig:40673-S5:a}
        \includegraphics[width=\textwidth]{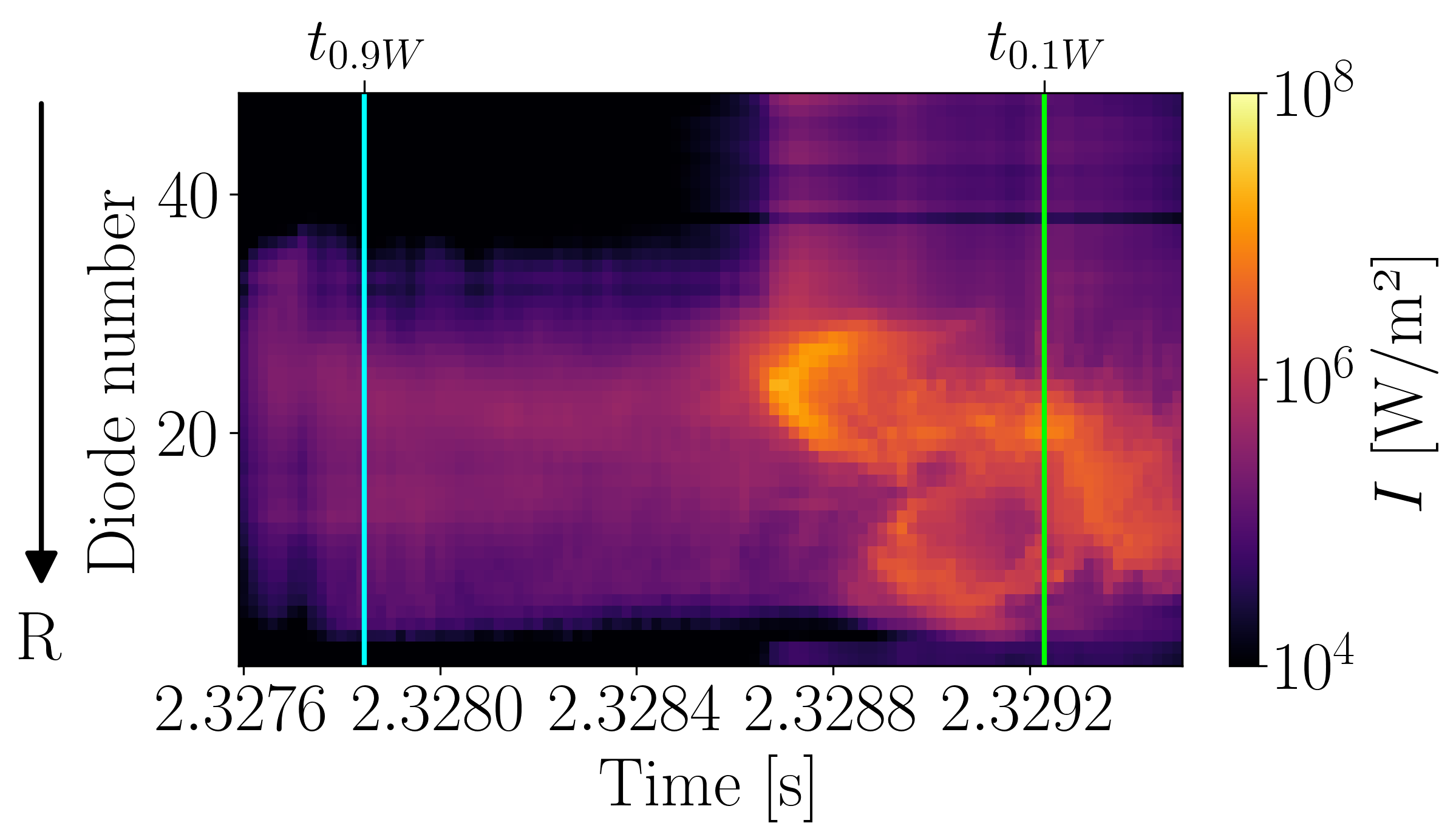}
    \end{subfigure}
    \hfill
    \begin{subfigure}[b]{0.49\textwidth}
        \centering
        \caption{Synthetic, sector 5, horizontal `DHC'}\label{fig:40673-S5:b}
        \includegraphics[width=\textwidth]{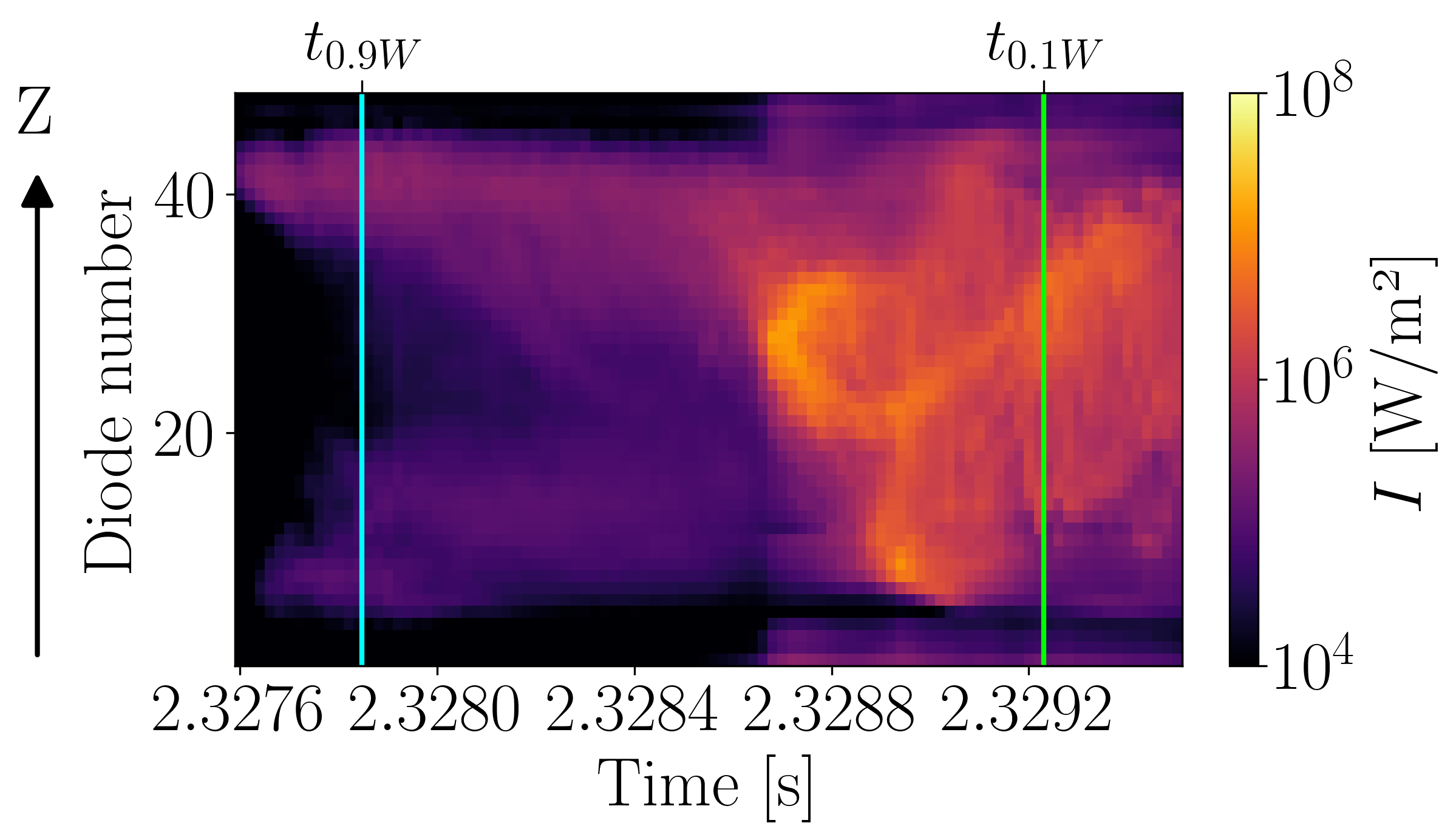}
    \end{subfigure}

    \begin{subfigure}[b]{0.49\textwidth}
        \centering
        \caption{Experimental, sector 5, vertical `DVC'}\label{fig:40673-S5:c}
        \includegraphics[width=\textwidth]{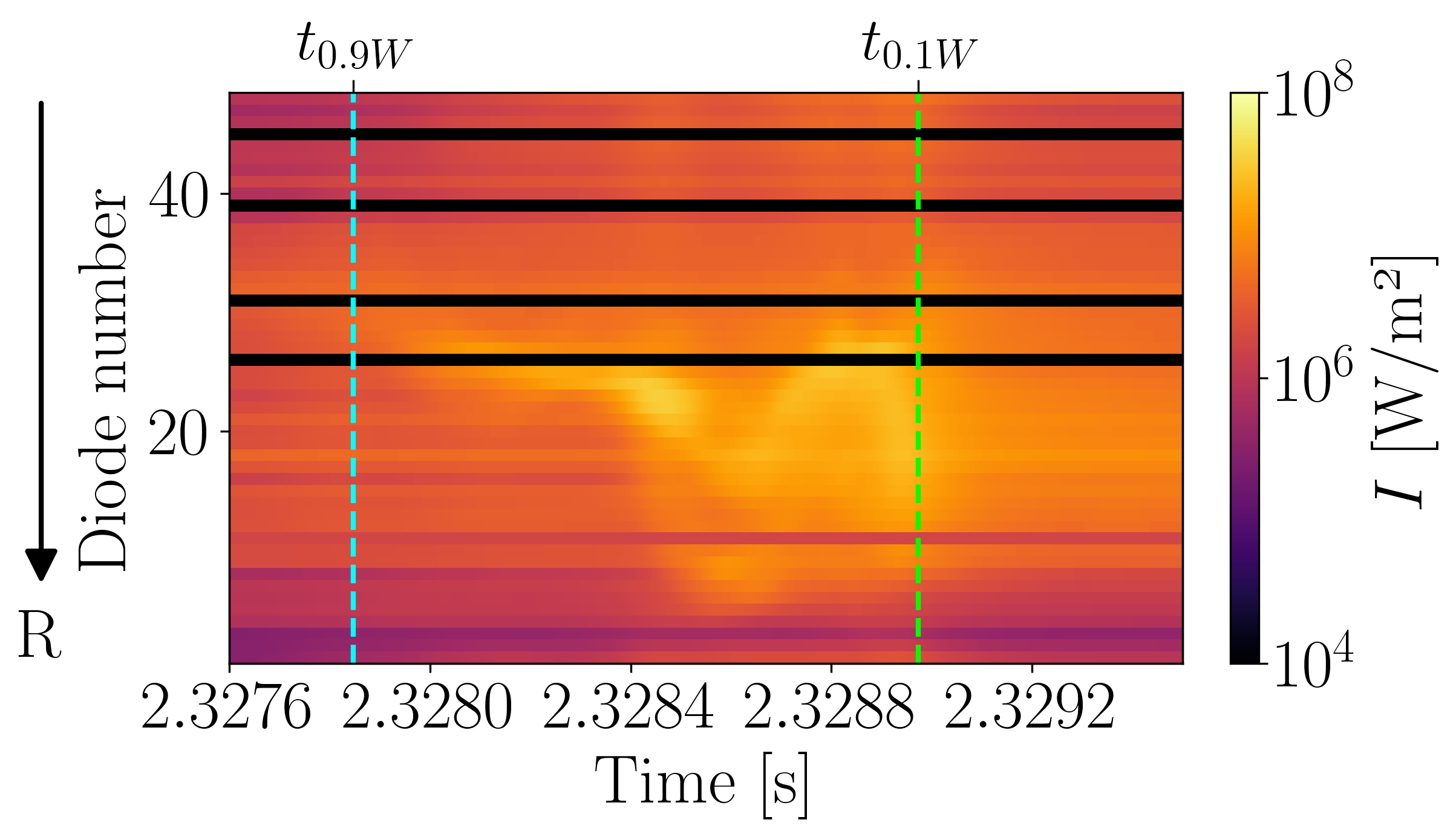}
    \end{subfigure}
    \hfill
    \begin{subfigure}[b]{0.49\textwidth}
        \centering
        \caption{Experimental, sector 5, horizontal `DHC'}\label{fig:40673-S5:d}
        \includegraphics[width=\textwidth]{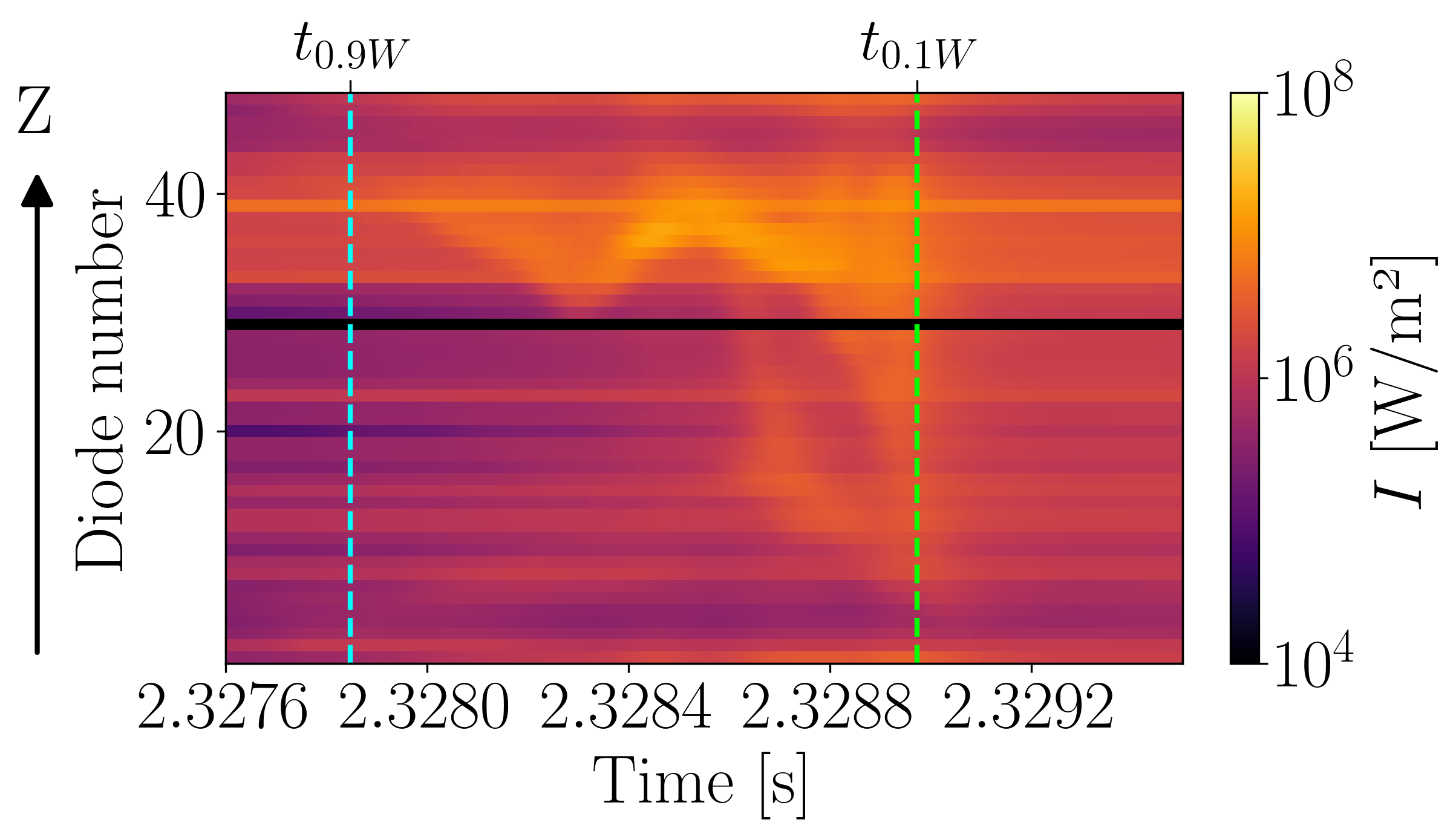}
    \end{subfigure}

    \caption{Comparison of synthetic and experimental AXUV data in sector 5 (110$^{\circ}$ away toroidally from the SPI) for AUG discharge \#40673.
    The shattered pellet contained 10\,mol\% Ne.
    Figures (a) and (b) show the results of the synthetic measurement performed on the JOREK data, while figures (c) and (d) present the experimental data.}\label{fig:40673-S5}
\end{figure}

The synthetic signals of the JOREK simulations of AUG discharge \#40673 are shown in the top plots of figures~\ref{fig:40673-S16} and \ref{fig:40673-S5} for sectors 16 and 5 of AUG, respectively.
The experimental signals are plotted just below the simulated signals to facilitate the comparison of the radiation features.
Note that the SPI is located in sector 16, and sector 5 is toroidally $\sim$110$^{\circ}$ apart from the injection location.
The figures are contour plots of logarithmically scaled AXUV line-integrated brightness signals, with the same units and scales for all synthetic and experimental figures.
The plotted time range in figures~\ref{fig:40673-S16}, \ref{fig:40673-S5} corresponds to the time range plotted in figure \ref{fig:40673_W_th}.
The starting times roughly correspond to the time of arrival of the first pellet fragments in both the simulations and the experiments.
The solid lines in the synthetic data figures and the dashed lines in the experimental data figures mark the times at which $W_{\mathrm{th}}(t_{0.9W})=0.9W_{\mathrm{th, 0}}$ and $W_{\mathrm{th}}(t_{0.1W})=0.1W_{\mathrm{th, 0}}$.

Comparing the vertical camera signals for discharge \#40673 with the 10\% Ne pellet, the figure shows that the initial inward penetration of the radiation cloud in the synthetic data (figure~\ref{fig:40673-S16:a}) and in the experimental data (figure~\ref{fig:40673-S16:c}) are in a good match in velocity and amplitude until around $t=2.3285$~s.
This confirms the expectation of a straightforward trajectory-matching related to the higher Ne-content in the pellet.
Additionally, fine radiation structures can be found in both figures, moving from the main radiation cloud towards the high-field side during this initial part of the discharge.
Together with the horizontal data in figures~\ref{fig:40673-S16:b} and \ref{fig:40673-S16:d}, this phenomenon suggests that the radiation spreads along the outer flux surfaces, vertically downward and towards the high-field side.
After $t=2.3285$~s, the synthetic data from the vertical camera show a single abrupt radial spreading of the radiation cloud.
In contrast, the experimental data continue to show multiple smaller events before and after this time.
The evolution of the spatial extent of the radiation cloud, however, appears consistent across the experimental and synthetic data in both the vertical and horizontal figures.

At this point, it is important to note that the shattering of the pellet during SPI experiments has a significant stochastic component, as do the resulting fragment size and velocity distributions, even though their statistics depend on the injection velocity and the shatter head geometry.
The effects of this are apparent in earlier works~\citep{tang_quantitative_2026,halldestam_reduced_2025}, where simulations with different fragment numbers are compared.
The fragmentation in these simulations follows the model of Parks~\citep{parks_modeling_2016,tang_non-linear_2025}, which is known to overestimate fragment sizes, and this may contribute to the differences observed here.
The exact fragment numbers and velocity distributions are also unknown in the experiments, and some of the differences in the simulated and measured AXUV data might stem from these phenomena.
Additionally, the JOREK simulations did not include background impurities.
Consequently, the background radiation in the synthetic data is due to reflections of radiation emitted by the pellet cloud, whereas in the experiment, we can expect some radiating impurities, such as tungsten, to be present in the plasma.
This effect is most apparent during the first few time steps.
However, because the pellet contains a relatively large amount of neon, its radiation dominates the background emission, making the effect less significant for this discharge.
The 3D AXUV synthetic diagnostic, in a sense, can validate not only the JOREK simulations themselves but also the underlying physics models used.

In the horizontal experimental camera data (figure \ref{fig:40673-S16:d}), we see an abrupt vertical spreading just before $t=2.3285$~s, whereas in the synthetic figure (figure \ref{fig:40673-S16:b}), it appears to occur a bit later and with greater intensity.
However, it is important to distinguish between radiation and material deposition, as material deposited in a poloidal location could radiate more intensely at some point after deposition, as the local temperature drops during the disruption.
Examining both datasets after the thermal energy reaches 20\% of its initial value, the fine radiation structures are quickly smoothed out and disappear, possibly signalling significant mixing.
Overall, the synthetic data qualitatively matches the experimental data well, and some features match quantitatively as well. This shows that the synthetic diagnostic is a capable tool for such comparisons.

Figure~\ref{fig:40673-S5} shows similar synthetic-experimental and vertical-horizontal comparisons of sector 5 AXUV cameras for the same \#40673 discharge.
Here, the effect of the background impurity radiation is more apparent than in sector 16, as these cameras are toroidally $\sim110^{\circ}$ away from the SPI location, and the toroidally spreading neon radiates in a less localised manner.
This is also evident in the synthetic signals in figures~\ref{fig:40673-S5:a} and~\ref{fig:40673-S5:b}, where there is no impurity radiation, only the injected neon radiates, and we see significantly less radiation than in the corresponding data in sector 16.
The ablated pellet material quickly ionises, elongates along the helical field lines on a short timescale, and therefore the location where it radiates in sector~5 differs from that in sector~16.
The field line following elongation has been confirmed in a previous experimental data analysis effort for the initial stage of the shattered pellet injection~\citep{lengyel_plasma_2024}.
Comparing the synthetic to the experimental data in the figure~\ref{fig:40673-S5:a}--\ref{fig:40673-S5:c} and \ref{fig:40673-S5:b}--\ref{fig:40673-S5:d} pairs, we can see larger differences.
In the experimental signals, the radiation cloud appears around the $t_{0.9W}$ mark, whereas in the simulation, the distinct radiation only appears closer to the $t_{0.1W}$ mark.
This delay in the onset of radiation might be due to the simulations by Tang~\citep{tang_quantitative_2026} not accounting for the equilibrium toroidal flow.
A similar radiation-line pattern is observed in the simulation, but with a much weaker amplitude. This persists until the core collapse is triggered by the $1/1$ mode~\citep{tang_quantitative_2026}, after which the radiation becomes more globally distributed and toroidally uniform.
Regardless of these differences, after the thermal energy drops to 20\% of its initial value, the radiation patterns begin to smooth out in both cases.

\begin{figure}
  \centering
  \includegraphics[width=0.75\textwidth]{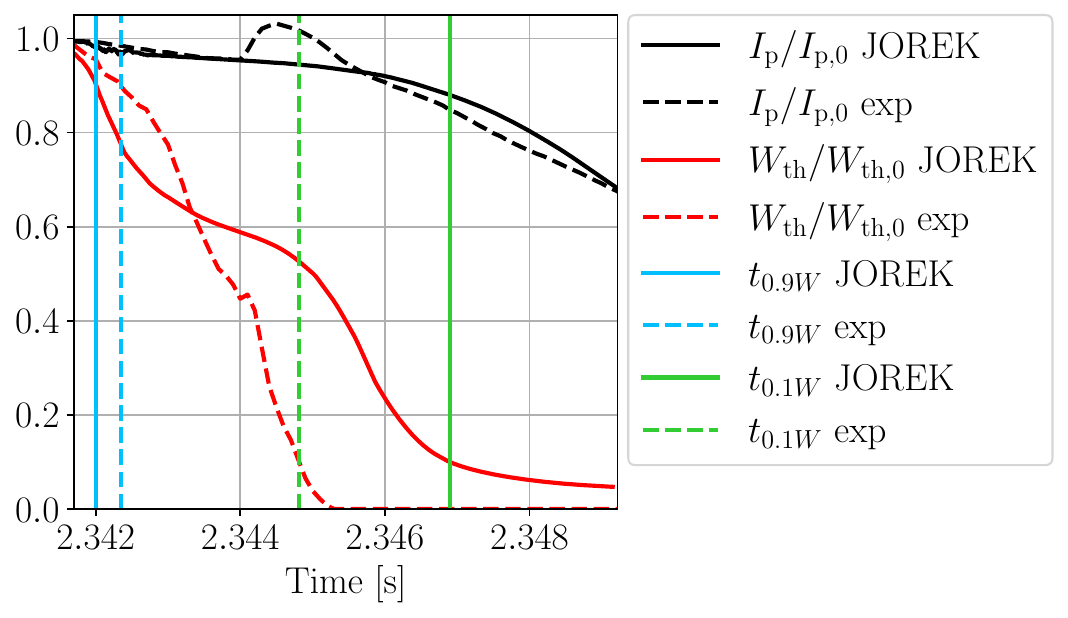}
  \caption{The time evolution of normalised plasma parameters from the AUG experiment \#41007 (0.17\% Ne), and of the corresponding JOREK simulation (based on figure 11 of Tang~\citep{tang_quantitative_2026}).
  Solid lines represent the simulation data, and dashed lines the experimental data.
  The black lines denote the normalised plasma current $I_{\mathrm{p}}$, and the red lines the normalised thermal energy $W_{\mathrm{th}}$.
  The 90\% and 10\% thermal energy points $t_{0.9W}$ and $t_{0.1W}$ are marked, with dashed lines for the experiment and solid lines for the simulation data.
  }\label{fig:41007_W_th}
\end{figure}

The second example shown in the paper concerns AUG discharge \#41007, in which the shattered pellet contained 0.17\,mol\% Ne~\citep{tang_quantitative_2026}.
The simulated and experimental time evolution of the normalised plasma current, thermal energy and radiated fraction are shown in figure~\ref{fig:41007_W_th}.
This is a relatively low neon content in the pellet, so lower radiation power and a longer disruption timescale are expected, and these are successfully reproduced in the simulation.

Due to the low neon content, the pellet is expected to be subject to the rocket effect, and its radiation cloud will drift~\citep{corbett_numerical_2026,vallhagen_drift_2023,Vallhagen_2025}.
Since the JOREK simulation did not account for these effects~\citep{tang_quantitative_2026}, the match in time evolution of the plotted plasma parameters is noticeably worse compared to the higher neon content case.
Even so, we may once again look to the $t_{0.9W}$ and $t_{0.1W}$ time points, since in this case, there is a more significant difference in the disruption's timescale between the simulation and the experiment, further motivating a physics-based comparison of the timescales.

\begin{figure}
    \centering
    \begin{subfigure}[b]{0.49\textwidth}
        \centering
        \caption{Synthetic, sector 16, vertical `D16'}\label{fig:41007-S16:a}
        \includegraphics[width=\textwidth]{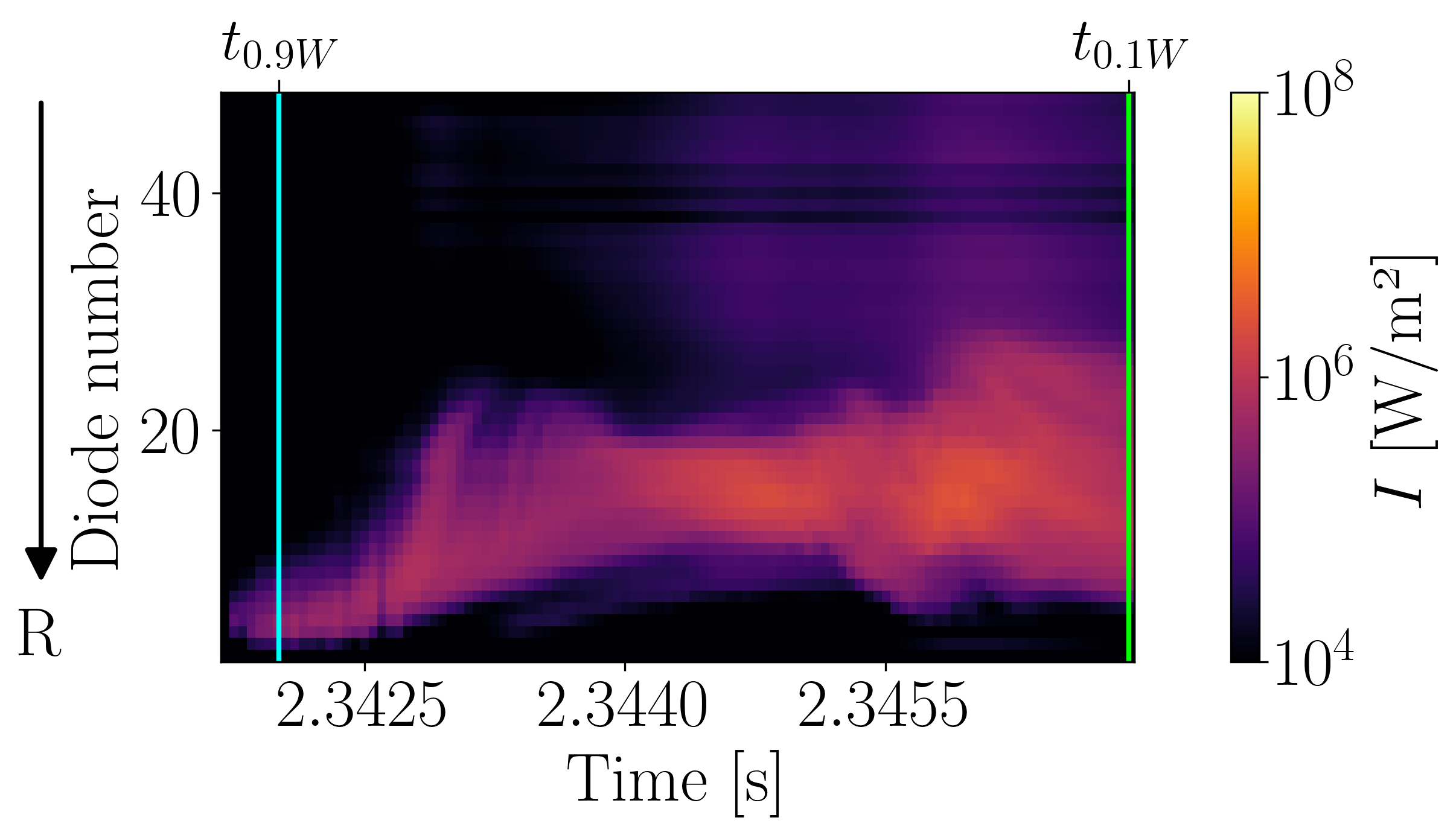}
    \end{subfigure}
    \hfill
    \begin{subfigure}[b]{0.49\textwidth}
        \centering
        \caption{Synthetic, sector 16, horizontal `DHT'}\label{fig:41007-S16:b}
        \includegraphics[width=\textwidth]{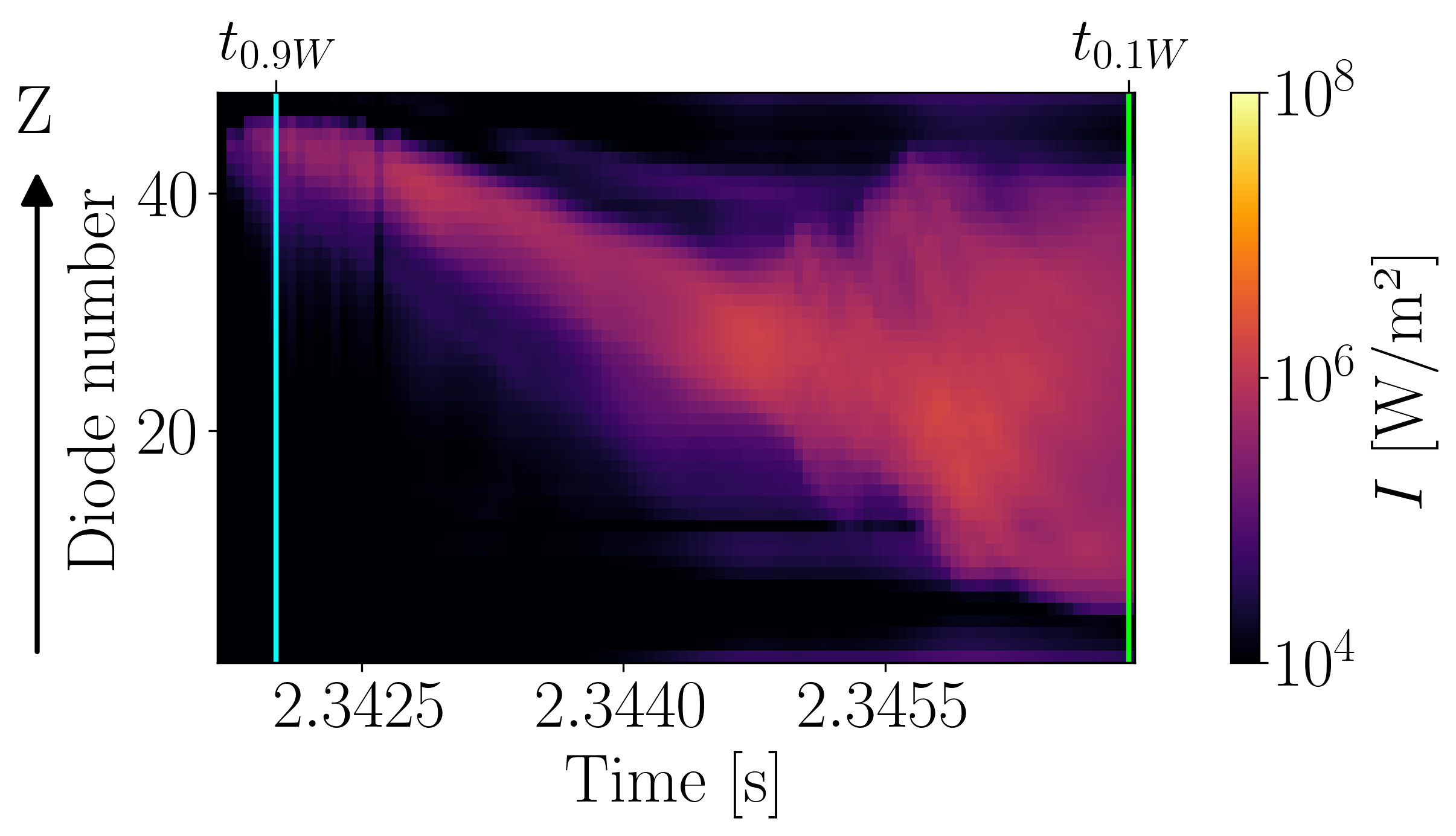}
    \end{subfigure}

    \begin{subfigure}[b]{0.49\textwidth}
        \centering
        \caption{Experimental, sector 16, vertical `D16'}\label{fig:41007-S16:c}
        \includegraphics[width=\textwidth]{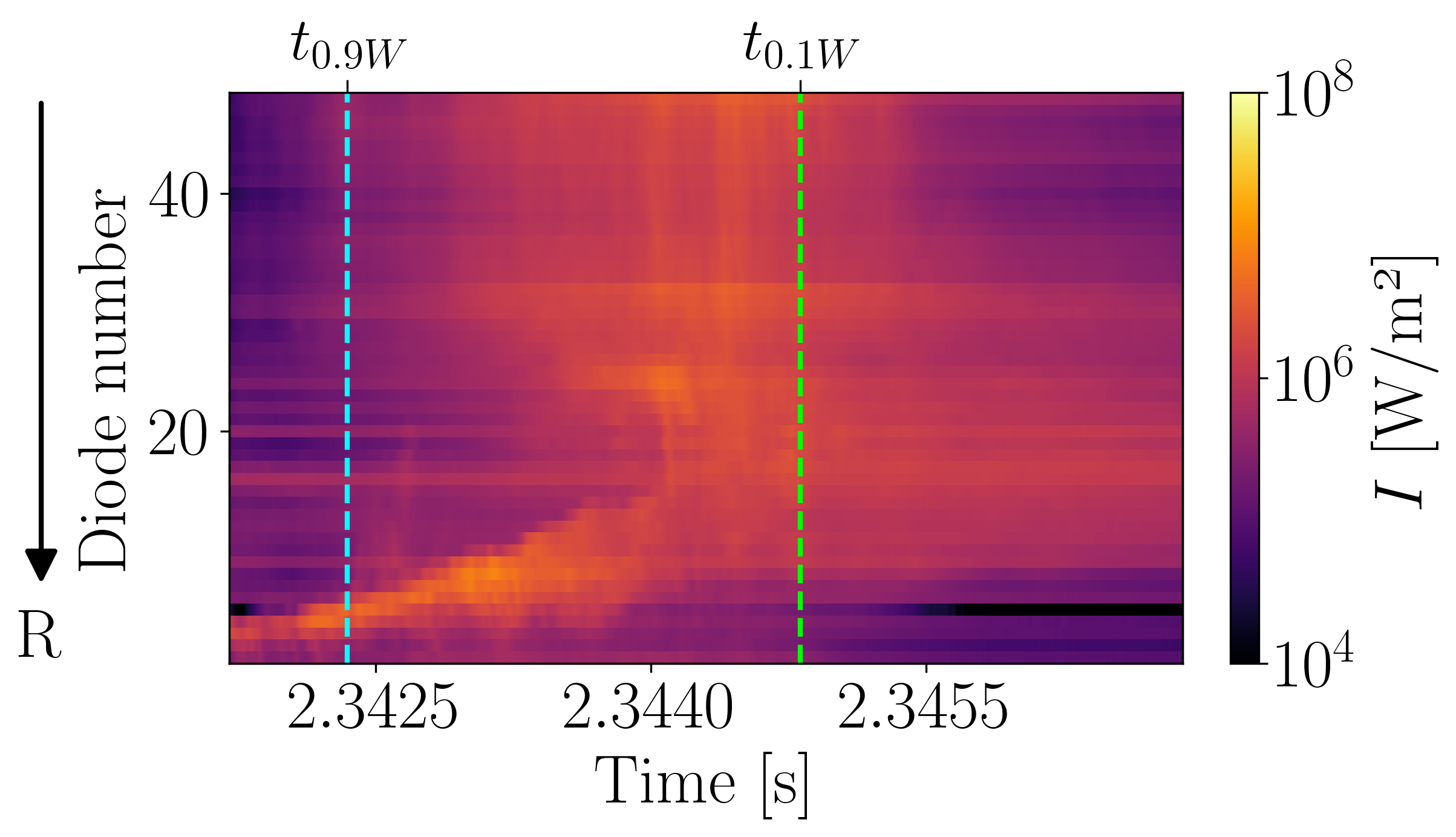}
    \end{subfigure}
    \hfill
    \begin{subfigure}[b]{0.49\textwidth}
        \centering
        \caption{Experimental, sector 16, horizontal `DHT'}\label{fig:41007-S16:d}
        \includegraphics[width=\textwidth]{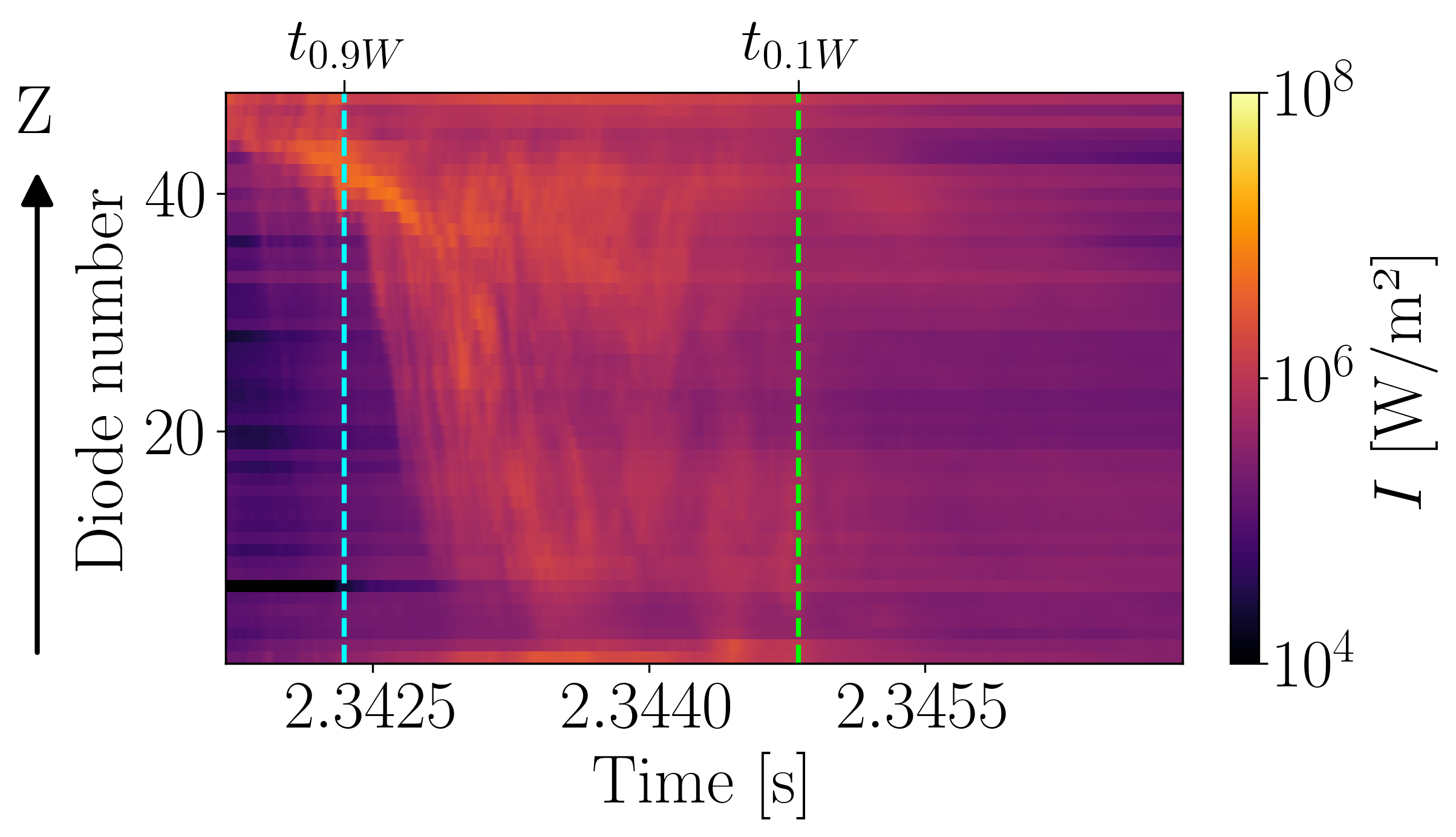}
    \end{subfigure}

    \caption{Comparison of synthetic and experimental AXUV data in sector 16 for AUG SPI discharge \#41007.
    The shattered pellet contained 0.17\,mol\% Ne.
    Figures (a) and (b) show the results of the synthetic measurement performed on the JOREK data, while figures (c) and (d) present the experimental data.}\label{fig:41007-S16}
\end{figure}

Figure~\ref{fig:41007-S16} presents the sector 16 comparison for discharge \#41007 with the synthetic figures in the top row and the experimental data in the bottom row.
The colour bar is the same as before, for the 10\,mol\% Ne pellet, so it is immediately clear that according to the expectations, we do indeed get significantly lower radiated power from the pellet with 0.17\,mol\% Ne.
The radiation from background impurities is not as suppressed by the radiation from the pellet as in the previously presented experiment.
As shown in the experimental measurements (figures~\ref{fig:41007-S16:c} and \ref{fig:41007-S16:d}), the background radiation is comparable in magnitude to the radiation from the pellet cloud.
This is in clear contrast to figures~\ref{fig:41007-S16:a} and \ref{fig:41007-S16:b}, since the simulation did not account for intrinsic impurities.
During the second half of the disruption in the synthetic AXUV signals, reflections appear, resembling the spatial spreading of the radiation.

The initial pellet trajectories still match in this comparison between experimental and synthetic data.
However, systematic discrepancies appear approximately 1~ms after fragment arrival.
These differences are more significant than those observed for a higher neon content pellet.
The differences likely stem from the aforementioned rocket effect and plasmoid drift in the experiment~\cite{Vallhagen_2025}, and this is nicely shown in the horizontal camera data in figure~\ref{fig:41007-S16:d}.
Very fine vertical radiation structures are already visible from the first plotted time step through around $t=2.3425$~s, likely corresponding to small fragments and their radiation clouds drifting downward.
After $t=2.3425$~s, a significant portion of the radiation cloud abruptly moves towards the bottom of the vacuum vessel, and in the remaining time until the 20\% thermal energy content point, a significant mixing occurs, which at some point around $t=2.344$~s is already visible in the vertical camera view too (figure~\ref{fig:41007-S16:c}).
Even though the fine radiation structures connected to pellet fragments and ablation clouds experiencing drifts are not visible in the synthetic data, if we look at the time point of $t_{0.1W}$, what the AXUV cameras see from the physics does match up with the experimental results: by this time, the radiation has fully spread out poloidally.
It just takes significantly more time in the simulated case, not including the rocket effect, than in the experiments.
Based on the present analysis, a recommendation is to use a pellet ablation model that includes the rocket effect to better reproduce material deposition and the thermal quench time in trace-Neon SPI scenarios.

Sector 5 comparisons for this discharge are omitted due to the very low visibility of any radiation structures in the experimental data and to retain focus on core findings.

\section{Conclusion}\label{sec:conclusion}
The AXUV measurement system on ASDEX Upgrade detects radiation across a wide spectral range with high time resolution, which makes it uniquely suited to studying the time evolution of the radiated power during shattered pellet injection (SPI).
Its measurements are, however, subject to significant uncertainty in dynamic scenarios.
Using a single line of sight and a spatially 1D DREAM simulation of AUG discharge \#40949~\citep{halldestam_reduced_2025}, we showed that the time-varying emission spectra during SPI make the effective response of the diodes evolve throughout the discharge.
Together with the unknown relative degradation between channels, this led us to abandon tomographic radiated power reconstruction~\citep{lengyel_plasma_2024} and to compare features in the measured and simulated signals instead.

A 3D synthetic diagnostic was then built on the Raysect ray-tracing framework~\citep{meakins_raysectsource_2023} and applied to two sectors of AUG, for a total of 192 lines of sight.
It employs the ray transfer method~\citep{kajita_usage_2016} over the 1\,eV to 1.24\,keV range in 300 spectral bins, and imports the full CAD geometry of the plasma-facing components to compute spectrally dependent reflections.
It was verified against analytically calculated etendue values (mean relative difference of 2.3\%) and cross-checked against the Cherab \texttt{PowerPipeline} Monte Carlo method (average relative error 0.017\%).
Validation against standard plasma discharges is hindered by the uncertainties in the tungsten impurity concentration and atomic data, so mixed Ne/D$_2$ SPI was chosen instead, where the neon radiation dominates the background plasma emission.
On this basis, even order-unity uncertainties in the relative channel sensitivities do not invalidate the comparisons.

Four results follow from applying the diagnostic to the JOREK simulations of Tang et al~\citep{tang_quantitative_2026}.
First, the radiation front can be tracked quantitatively for discharge \#40673 the inward radial penetration of the radiation cloud matches the experiment in both velocity and amplitude until $t=2.3285$~s.
The radiation front can also be tracked at the very beginning for discharge \#41007, albeit the lack of background impuritries hinders the quantitative comparison.
Second, the field line following elongation of the ablated material is directly observable $\sim$110$^{\circ}$ away from the injection, in both the synthetic and the experimental signals.
Third, the absence of a physics model in a simulation is detectable in the AXUV signal.
For discharge \#41007 (0.17\,mol\% Ne) the experiment shows fine vertical radiation structures drifting towards the bottom of the vessel, which are absent from the simulation because it does not account for the rocket effect or plasmoid drift~\citep{tang_quantitative_2026}.
The radiation nevertheless spreads across the poloidal cross-section by $W_\mathrm{th}=0.1\,W_\mathrm{th,0}$ in both cases.
Fourth, reflections from the plasma-facing components set the background level for practically all diodes once the radiation is strong, as no intrinsic impurities were included in the simulations.

The synthetic diagnostic is a capable tool for comparing simulated and measured AXUV signals, and it can be used to assess which physics models a simulation still needs.
For the cases studied here, it points to the inclusion of the rocket effect and plasmoid drift, background impurities, and toroidal equilibrium flows in future simulations.

\ack{This work has been carried out within the framework of the EUROfusion Consortium, funded by the European Union via the Euratom Research and Training Programme (Grant Agreement No 101052200 — EUROfusion). Views and opinions expressed are however those of the author(s) only and do not necessarily reflect those of the European Union or the European Commission. Neither the European Union nor the European Commission can be held responsible for them. The ASDEX Upgrade SPI project has been implemented as part of the ITER DMS Task Force programme. The SPI system and related diagnostics have received funding from the ITER Organization. The views and opinions expressed herein do not necessarily reflect those of the ITER Organization. F. Lengyel acknowledges the support of the JNEA Foundation.}

\suppdata{Movie 1. (0.5 MB MP4) Time evolution of the emitted radiated power density in the poloidal cross-section for the JOREK simulation of AUG discharge \#40673 (10\,mol\% Ne / 90\,mol\% D$_2$); a single frame is shown in figure~\ref{fig:example_emission}.\\*
Movie 2. (0.4 MB MP4) Time evolution of the emitted radiated power density in the poloidal cross-section for the JOREK simulation of AUG discharge \#41007 (0.17\,mol\% Ne).}

\bibliographystyle{iopart-num}
\bibliography{bibliography}

@article{Barabaschi_2026,
doi = {10.1088/1741-4326/ae4d5c},
url = {https://doi.org/10.1088/1741-4326/ae4d5c},
year = {2026},
month = {may},
publisher = {IOP Publishing},
volume = {66},
number = {11},
pages = {116001},
author = {Barabaschi, Pietro and Artola, Javier and Oliva, Alessandro Bonito and Carannante, Giuseppe and Coblentz, Laban and Encheva, Anna and Giniiatulin, Radmir and Grillot, David and Hunt, Ryan and Jachmich, Stefan and Kamada, Yutaka and Kim, Sun-Hee and Loarte, Alberto and Marquez, Alfonso and Merola, Mario and Noh, Chang Hyun and Nunes, Isabel and Orlandi, Sergio and Perrier, Gilles and Pinches, Simon and Pitts, Richard and Poli, Francesca and Reich, Jens and Schild, Thierry and Schneider, Mireille and Vaghela, Hiten and Veltri, Pierluigi},
title = {Progress of ITER and its importance for fusion development},
journal = {Nuclear Fusion}
}

@article{bernert_application_2014,
	title = {Application of {AXUV} diode detectors at {ASDEX} {Upgrade}},
	volume = {85},
	issn = {0034-6748},
	url = {https://doi.org/10.1063/1.4867662},
	doi = {10.1063/1.4867662},
	number = {3},
	urldate = {2025-03-10},
	journal = {Review of Scientific Instruments},
	author = {Bernert, M. and Eich, T. and Burckhart, A. and Fuchs, J. C. and Giannone, L. and Kallenbach, A. and McDermott, R. M. and Sieglin, B. and {ASDEX Upgrade Team}},
	month = mar,
	year = {2014},
	pages = {033503}
}

@article{Bodner_2025,
doi = {10.1088/1741-4326/add170},
url = {https://doi.org/10.1088/1741-4326/add170},
year = {2025},
month = {may},
publisher = {IOP Publishing},
volume = {65},
number = {6},
pages = {066010},
author = {Bodner, G. and Eidietis, N. and Chen, Z. and Heinrich, P. and Herfindal, J. and Jachmich, S. and Papp, G. and Kim, J. and Lehnen, M. and Sheikh, U. and Coffey, I. and Ficker, O. and Gerasimov, S. and Kachkanov, V. and Reux, C. and Silburn, S. and Sun, H. and the ASDEX Upgrade Team and JET Contributors and the EUROfusion Tokamak Exploitation Team},
title = {Multi-device analysis of energy loss duration and pellet penetration with implications for shattered pellet injection in ITER},
journal = {Nuclear Fusion}
}

@article{tang_non-linear_2025,
    title = {Non-linear {MHD} modeling of shattered pellet injection in {ASDEX} {Upgrade}},
    volume = {65},
    issn = {0029-5515},
    url = {https://doi.org/10.1088/1741-4326/adfce6},
    doi = {10.1088/1741-4326/adfce6},
    language = {en},
    number = {11},
    urldate = {2026-02-17},
    journal = {Nuclear Fusion},
    author = {Tang, W. and others},
    month = sep,
    year = {2025},
    note = {Publisher: IOP Publishing},
    pages = {116003},
}

@article{Heinrich_2024_SPI_Lab,
title = {Recipes for pellet generation and launching in the {ASDEX} {U}pgrade {SPI}},
journal = {Fusion Engineering and Design},
volume = {206},
pages = {114576},
year = {2024},
issn = {0920-3796},
doi = {10.1016/j.fusengdes.2024.114576},
url = {https://doi.org/10.1016/j.fusengdes.2024.114576},
author = {P. Heinrich and G. Papp and P. {de Marn\'e} and M. Dibon and S. Jachmich and M. Lehnen and T. Peherstorfer and I. Vinyar}
}

@article{heinrich_radiated_2025,
    title = {Radiated energy fraction of {SPI}-induced disruptions at {ASDEX} {Upgrade}},
    volume = {65},
    issn = {0029-5515},
    url = {https://doi.org/10.1088/1741-4326/adcbc0},
    doi = {10.1088/1741-4326/adcbc0},
    language = {en},
    number = {5},
    urldate = {2026-02-17},
    journal = {Nuclear Fusion},
    publisher = {IOP Publishing},
    author = {Heinrich, P. and others},
    month = apr,
    year = {2025},
    pages = {056036},
}

@phdthesis{Heinrich_2025_PhD,
	author = {Heinrich, Paul},
	title = {Shattered pellet injection studies at the tokamak {ASDEX Upgrade}},
	year = {2025},
	school = {Technische Universit\"at M\"unchen},
	pages = {154},
	language = {en},
	url = {https://mediatum.ub.tum.de/1767878}
}

@misc{Heinrich_2026_disr_evo,
	title={Evolution of {SPI}-induced disruptions in {ASDEX Upgrade}},
	author={Heinrich, P. and Papp, G. and Jachmich, S. and Artola, J. and Bernert, M. and {de~Marn\'e}, P. and Dibon, M. and Dux, R. and Eberl, T. and Halldestam, P. and Hobirk, J. and Hoelzl, M. and Klossek, F. and Lehnen, M. and Lunt, T. and Maraschek, M. Patel, A. and Peherstorfer, T. and Schwarz, N. and Sheikh, U. and Sieglin, B. and Svoboda, J. and Tang, W. and {the ASDEX Upgrade Team} and {the EUROfusion Tokamak Exploitation Team}},
	year={2026},
	publisher = {IOP Publishing},
	journal = {submitted to Nuclear Fusion, pre-print available at \url{https://doi.org/10.48550/arXiv.2604.05488}},
	note = {submitted to Nuclear Fusion, pre-print available at \url{https://doi.org/10.48550/arXiv.2604.05488}},
	eprint={2604.05488},
      	archivePrefix={arXiv},
}

@inproceedings{carr_towards_2017,
    address = {Belfast, Northern Ireland (UK)},
    series = {Europhysics {Conference} {Abstracts}},
    booktitle = {44th EPS Plasma Physics Conference},
    title = {Towards integrated data analysis of divertor diagnostics with ray-tracing},
    volume = {41F},
    isbn = {979-10-96389-07},
    language = {en},
    author = {Carr, M and others},
    year = {2017},
}

@misc{carr_cherabcore_2023,
    title = {cherab/core: {Release} 1.4.0},
    shorttitle = {cherab/core},
    url = {https://zenodo.org/records/7603688},
    doi = {10.5281/zenodo.7603688},
    urldate = {2026-02-17},
    publisher = {Zenodo},
    author = {Carr, Matthew and Lovell, Jack and Tomes, Matej and Neverov, Vlad and Meakins, Dr Alex and MUNECHIKA, Koyo and Stańczak-Marikin, Dominik and Bold, David},
    month = feb,
    year = {2023},
}

@article{dibon_design_2023,
    title = {Design of the shattered pellet injection system for {ASDEX} {Upgrade}},
    volume = {94},
    issn = {0034-6748},
    url = {https://doi.org/10.1063/5.0141799},
    doi = {10.1063/5.0141799},
    number = {4},
    urldate = {2025-04-14},
    journal = {Review of Scientific Instruments},
    author = {Dibon, M. and others},
    month = apr,
    year = {2023},
    pages = {043504},
}

@phdthesis{patel_modelling_2023,
    title = {Modelling of shattered pellet injection experiments on the {ASDEX} {Upgrade} tokamak},
    url = {http://arxiv.org/abs/2312.03462},
    doi = {10.48550/arXiv.2312.03462},
    urldate = {2026-02-19},
    school = {Eindhoven University of Technology},
    author = {Patel, Anshkumar Himanshu},
    month = dec,
    year = {2023},
    note = {arXiv:2312.03462 [physics]},
}

@inproceedings{lengyel_plasma_2024,
    address = {Salamanca, Spain},
    series = {Europhysics {Conference} {Abstracts}},
    title = {Plasma radiation characteristics of {SPI} mitigated disruptions in {ASDEX} {Upgrade}},
    volume = {48A},
    isbn = {111-22-33333-44-5},
    url = {https://lac913.epfl.ch/epsppd3/2024/html/PDF/P5-041.pdf},
    language = {en},
    booktitle = {50th {EPS} {Conference} on {Plasma} {Physics}},
    author = {Lengyel, F and others},
    year = {2024},
}

@article{hoelzl_jorek_2021,
    title = {The {JOREK} non-linear extended {MHD} code and applications to large-scale instabilities and their control in magnetically confined fusion plasmas},
    volume = {61},
    issn = {0029-5515},
    url = {https://dx.doi.org/10.1088/1741-4326/abf99f},
    doi = {10.1088/1741-4326/abf99f},
    language = {en},
    number = {6},
    urldate = {2025-04-06},
    journal = {Nuclear Fusion},
    publisher = {IOP Publishing},
    author = {Hoelzl, M. and others},
    month = may,
    year = {2021},
    pages = {065001},
}

@phdthesis{tal_measurement_2015,
    address = {Budapest, Hungary},
    type = {{PhD}},
    title = {Measurement of transient events in hot magnetized plasmas},
    url = {https://repozitorium.omikk.bme.hu//bitstream/handle/10890/1448/ertekezes.pdf},
    language = {en},
    urldate = {2026-02-24},
    school = {Budapest University of Technology and Economics},
    author = {T{\'a}l, B.},
    year = {2015},
}

@article{canfield_stability_1989,
    title = {Stability and quantum efficiency performance of silicon photodiode detectors in the far ultraviolet},
    volume = {28},
    copyright = {https://doi.org/10.1364/OA\_License\_v1\#VOR},
    issn = {0003-6935, 1539-4522},
    url = {https://opg.optica.org/abstract.cfm?URI=ao-28-18-3940},
    doi = {10.1364/AO.28.003940},
    language = {en},
    number = {18},
    urldate = {2026-02-19},
    journal = {Applied Optics},
    author = {Canfield, L. R. and Kerner, J. and Korde, R.},
    month = sep,
    year = {1989},
    pages = {3940},
}

@article{korde_quantum_1987,
    title = {Quantum efficiency stability of silicon photodiodes},
    volume = {26},
    copyright = {https://doi.org/10.1364/OA\_License\_v1\#VOR},
    issn = {0003-6935, 1539-4522},
    url = {https://opg.optica.org/abstract.cfm?URI=ao-26-24-5284},
    doi = {10.1364/AO.26.005284},
    language = {en},
    number = {24},
    urldate = {2026-02-19},
    journal = {Applied Optics},
    author = {Korde, R. and Geist, J.},
    month = dec,
    year = {1987},
    pages = {5284},
}

@misc{meakins_raysectsource_2023,
    title = {raysect/source: v0.8.1 {Release}},
    shorttitle = {raysect/source},
    url = {https://zenodo.org/records/7633656},
    doi = {10.5281/zenodo.7633656},
    urldate = {2026-03-19},
    publisher = {Zenodo},
    author = {Meakins, A. and Carr, M. and Orchard, S. and Lovell, J. and Neverov, V. and Munechika, K. and Tomes, M. and Essen, M. von},
    month = feb,
    year = {2023},
}

@article{halldestam_reduced_2025,
    title = {Reduced kinetic modelling of shattered pellet injection in {ASDEX} upgrade},
    volume = {91},
    issn = {0022-3778, 1469-7807},
    url = {https://www.cambridge.org/core/journals/journal-of-plasma-physics/article/reduced-kinetic-modelling-of-shattered-pellet-injection-in-asdex-upgrade/6CAE9C8A9A2676D38D1F6230029E254E},
    doi = {10.1017/S0022377825100470},
    language = {en},
    number = {4},
    urldate = {2026-03-23},
    journal = {Journal of Plasma Physics},
    author = {Halldestam, Peter and Heinrich, Paul and Papp, Gergely and Hoppe, Mathias and Hoelzl, Matthias and Pusztai, Istvan and Vallhagen, Oskar and Fischer, Rainer and Jenko, Frank and {ASDEX Upgrade Team} and {EUROfusion Tokamak Exploitation Team}},
    month = aug,
    year = {2025},
    pages = {E104},
}

@article{hoppe_dream_2021,
    title = {{DREAM}: {A} fluid-kinetic framework for tokamak disruption runaway electron simulations},
    volume = {268},
    issn = {0010-4655},
    shorttitle = {{DREAM}},
    url = {https://www.sciencedirect.com/science/article/pii/S0010465521002101},
    doi = {10.1016/j.cpc.2021.108098},
    urldate = {2025-02-18},
    journal = {Computer Physics Communications},
    author = {Hoppe, Mathias and Embreus, Ola and Fülöp, Tünde},
    month = nov,
    year = {2021},
    pages = {108098},
}

@misc{summers_adas_2004,
    title = {The {ADAS} {User} {Manual}, version 2.6},
    url = {https://www.adas.ac.uk},
    urldate = {2026-03-23},
    author = {Summers, H. P.},
    year = {2004},
}

@article{tang_quantitative_2026,
	author={Tang, W. and others},
	title={{Direct comparison of 3D non-linear JOREK simulations of shattered pellet injection with ASDEX Upgrade experiments}},
	journal={Nuclear Fusion},
	url={http://iopscience.iop.org/article/10.1088/1741-4326/ae9966},
	year={2026},
    volume = {},
    pages = {in press}
}

@article{windt_optical_1988,
    title = {Optical constants for thin films of {Ti}, {Zr}, {Nb}, {Mo}, {Ru}, {Rh}, {Pd}, {Ag}, {Hf}, {Ta}, {W}, {Re}, {Ir}, {Os}, {Pt}, and {Au} from 24 Å to 1216 Å},
    volume = {27},
    copyright = {© 1988 Optical Society of America},
    issn = {2155-3165},
    url = {https://opg.optica.org/ao/abstract.cfm?uri=ao-27-2-246},
    doi = {10.1364/AO.27.000246},
    language = {EN},
    number = {2},
    urldate = {2026-04-23},
    journal = {Applied Optics},
    publisher = {Optica Publishing Group},
    author = {Windt, David L. and Cash, Webster C. and Scott, M. and Arendt, P. and Newnam, B. and Fisher, R. F. and Swartzlander, A. B.},
    month = jan,
    year = {1988},
    pages = {246--278},
}

@article{angioni_impact_2025,
    title = {The impact of tungsten and tungsten transport on {H}-mode plasmas, experiments and modeling},
    volume = {65},
    issn = {0029-5515},
    url = {https://doi.org/10.1088/1741-4326/add1ee},
    doi = {10.1088/1741-4326/add1ee},
    language = {en},
    number = {6},
    urldate = {2026-05-07},
    journal = {Nuclear Fusion},
    publisher = {IOP Publishing},
    author = {Angioni, C.},
    month = may,
    year = {2025},
    pages = {062001},
}

@article{putterich_calculation_2010,
    title = {Calculation and experimental test of the cooling factor of tungsten},
    volume = {50},
    issn = {0029-5515},
    url = {https://doi.org/10.1088/0029-5515/50/2/025012},
    doi = {10.1088/0029-5515/50/2/025012},
    language = {en},
    number = {2},
    urldate = {2026-05-07},
    journal = {Nuclear Fusion},
    author = {Pütterich, T. and Neu, R. and Dux, R. and Whiteford, A.D. and O'Mullane, M.G. and Summers, H.P. and {ASDEX Upgrade Team}},
    month = jan,
    year = {2010},
    pages = {025012},
}

@article{peyrusse_tungsten_2026,
    title = {Tungsten ionization and emissivity in tokamak plasma conditions from the {Configuration} {Average} / {Unresolved} {Transition} {Array} approach},
    volume = {33},
    issn = {1070-664X},
    url = {https://doi.org/10.1063/5.0307327},
    doi = {10.1063/5.0307327},
    number = {1},
    urldate = {2026-07-10},
    journal = {Physics of Plasmas},
    publisher = {American Institute of Physics},
    author = {Peyrusse, Olivier and Desgranges, Corinne and Forestier-Colleoni, Pierre and Guirlet, Rémy},
    month = jan,
    year = {2026},
    pages = {013302},
}

@misc{corbett_numerical_2026,
    title = {Numerical model for pellet rocket acceleration in {PELOTON}},
    url = {http://arxiv.org/abs/2512.04484},
    doi = {10.48550/arXiv.2512.04484},
    urldate = {2026-05-07},
    publisher = {arXiv},
    author = {Corbett, J. and Samulyak, R. and Artola, F. J. and Jachmich, S. and Kong, M. and Nardon, E.},
    month = feb,
    year = {2026},
    note = {arXiv:2512.04484 [physics]},
}

@article{vallhagen_drift_2023,
    title = {Drift of ablated material after pellet injection in a tokamak},
    volume = {89},
    issn = {0022-3778, 1469-7807},
    url = {https://www.cambridge.org/core/journals/journal-of-plasma-physics/article/drift-of-ablated-material-after-pellet-injection-in-a-tokamak/404DC07A0F1E158CC6620E991B6D35B6},
    doi = {10.1017/S0022377823000466},
    language = {en},
    number = {3},
    urldate = {2026-05-07},
    journal = {Journal of Plasma Physics},
    author = {Vallhagen, O. and Pusztai, I. and Helander, P. and Newton, S. L. and Fülöp, T.},
    month = jun,
    year = {2023},
    pages = {905890306},
}

@article{polyanskiy_refractiveindexinfo_2024,
    title = {Refractiveindex.info database of optical constants},
    volume = {11},
    copyright = {2024 The Author(s)},
    issn = {2052-4463},
    url = {https://www.nature.com/articles/s41597-023-02898-2},
    doi = {10.1038/s41597-023-02898-2},
    language = {en},
    number = {1},
    urldate = {2026-05-07},
    journal = {Scientific Data},
    publisher = {Nature Publishing Group},
    author = {Polyanskiy, Mikhail N.},
    month = jan,
    year = {2024},
    pages = {94},
}

@article{matsuyama_neutral_2022,
    title = {Neutral gas and plasma shielding ({NGPS}) model and cross-field motion of ablated material for hydrogen–neon mixed pellet injection},
    volume = {29},
    issn = {1070-664X},
    url = {https://doi.org/10.1063/5.0084586},
    doi = {10.1063/5.0084586},
    number = {4},
    urldate = {2026-05-07},
    journal = {Physics of Plasmas},
    author = {Matsuyama, Akinobu},
    month = apr,
    year = {2022},
    pages = {042501},
}

@misc{lengyel_plasma-sdsbolometeraxuv_2026,
    title = {plasma-sds/bolometer/axuv: {AXUV} synthetic diagnostics for {ASDEX} {Upgrade}},
    copyright = {MIT license},
    url = {https://github.com/plasma-sds/bolometer},
    author = {Lengyel, Ferenc},
    month = may,
    year = {2026},
}

@article{kajita_usage_2016,
    title = {Usage of {Ray} {Tracing} {Transfer} {Matrix} to {Mitigate} the {Stray} {Light} for {ITER} {Spectroscopy}},
    volume = {56},
    copyright = {Copyright © 2016 WILEY-VCH Verlag GmbH \& Co. KGaA, Weinheim},
    issn = {1521-3986},
    url = {https://onlinelibrary.wiley.com/doi/abs/10.1002/ctpp.201500124},
    doi = {10.1002/ctpp.201500124},
    language = {en},
    number = {9},
    urldate = {2026-05-18},
    journal = {Contributions to Plasma Physics},
    author = {Kajita, S. and Veshchev, E. and Barnsley, R. and Walsh, M.},
    year = {2016},
    note = {\_eprint: https://onlinelibrary.wiley.com/doi/pdf/10.1002/ctpp.201500124},
    pages = {837--845},
}

@article{Pegourie_2007,
    doi = {10.1088/0741-3335/49/8/R01},
    url = {https://dx.doi.org/10.1088/0741-3335/49/8/R01},
    year = {2007},
    month = {jul},
    publisher = {},
    volume = {49},
    number = {8},
    pages = {R87},
    author = {B P{\'e}gouri{\'e}},
    title = {Review: Pellet injection experiments and modelling},
    journal = {Plasma Phys. Control. Fusion}
}

@article{kong_2026_plasmoid_drift,
	author={Kong, Mengdi and others},
	title={Role of plasmoid drift in the efficiency and reliability of shattered pellet injection},
	journal={Plasma Physics and Controlled Fusion},
	url={http://iopscience.iop.org/article/10.1088/1361-6587/ae72c7},
	year={2026}
}

@article{carr_description_2018,
    title = {Description of complex viewing geometries of fusion tomography diagnostics by ray-tracing},
    volume = {89},
    issn = {0034-6748},
    url = {https://doi.org/10.1063/1.5031087},
    doi = {10.1063/1.5031087},
    number = {8},
    urldate = {2026-06-17},
    journal = {Review of Scientific Instruments},
    author = {Carr, M. and Meakins, A. and Bernert, M. and David, P. and Giroud, C. and Harrison, J. and Henderson, S. and Lipschultz, B. and Reimold, F. and {EUROfusion MST1 Team} and {ASDEX Upgrade Team}},
    month = aug,
    year = {2018},
    pages = {083506},
}

@article{rakic_optical_1998,
    title = {Optical properties of metallic films for vertical-cavity optoelectronic devices},
    volume = {37},
    copyright = {© 1998 Optical Society of America},
    issn = {2155-3165},
    url = {https://opg.optica.org/ao/abstract.cfm?uri=ao-37-22-5271},
    doi = {10.1364/AO.37.005271},
    language = {EN},
    number = {22},
    urldate = {2026-06-29},
    journal = {Applied Optics},
    publisher = {Optica Publishing Group},
    author = {Rakić, Aleksandar D. and Djurišić, Aleksandra B. and Elazar, Jovan M. and Majewski, Marian L.},
    month = aug,
    year = {1998},
    pages = {5271--5283},
}

@article{lehnen_disruptions_2015,
    series = {{PLASMA}-{SURFACE} {INTERACTIONS} 21},
    title = {Disruptions in {ITER} and strategies for their control and mitigation},
    volume = {463},
    issn = {0022-3115},
    url = {https://www.sciencedirect.com/science/article/pii/S0022311514007594},
    doi = {10.1016/j.jnucmat.2014.10.075},
    urldate = {2026-07-08},
    journal = {Journal of Nuclear Materials},
    author = {Lehnen, M. and others},
    month = aug,
    year = {2015},
    pages = {39--48},
}

@article{loarte_new_2025,
    title = {The new {ITER} baseline, research plan and open {R}\&{D} issues},
    volume = {67},
    issn = {0741-3335},
    url = {https://doi.org/10.1088/1361-6587/add9c9},
    doi = {10.1088/1361-6587/add9c9},
    language = {en},
    number = {6},
    urldate = {2026-07-08},
    journal = {Plasma Physics and Controlled Fusion},
    publisher = {IOP Publishing},
    author = {Loarte, A and others},
    month = jun,
    year = {2025},
    pages = {065023},
}

@article{jachmich_shattered_2021,
    title = {Shattered pellet injection experiments at {JET} in support of the {ITER} disruption mitigation system design},
    volume = {62},
    issn = {0029-5515},
    url = {https://doi.org/10.1088/1741-4326/ac3c86},
    doi = {10.1088/1741-4326/ac3c86},
    language = {en},
    number = {2},
    urldate = {2026-07-15},
    journal = {Nuclear Fusion},
    publisher = {IOP Publishing},
    author = {Jachmich, S. and Kruezi, U. and Lehnen, M. and Baruzzo, M. and Baylor, L.R. and Carnevale, D. and Craven, D. and Eidietis, N.W. and Ficker, O. and Gebhart, T.E. and Gerasimov, S. and Herfindal, J.L. and Hollmann, E. and Huber, A. and Lomas, P. and Lovell, J. and Manzanares, A. and Maslov, M. and Mlynar, J. and Pautasso, G. and Paz-Soldan, C. and Peacock, A. and Piron, L. and Plyusnin, V. and Reinke, M. and Reux, C. and Rimini, F. and Sheikh, U. and Shiraki, D. and Silburn, S. and Sweeney, R. and Wilson, J. and Carvalho, P. and JET Contributors, the},
    month = dec,
    year = {2021},
    pages = {026012},
}

@article{lee_peridynamic_2024,
    title = {Peridynamic modelling of cryogenic deuterium pellet fragmentation for shattered pellet injection in tokamaks},
    volume = {64},
    issn = {0029-5515},
    url = {https://doi.org/10.1088/1741-4326/ad69a3},
    doi = {10.1088/1741-4326/ad69a3},
    language = {en},
    number = {10},
    urldate = {2026-07-16},
    journal = {Nuclear Fusion},
    publisher = {IOP Publishing},
    author = {Lee, S.-J. and Madenci, E. and Na, Yong-Su and de Marné, P. and Dibon, M. and Heinrich, P. and Jachmich, S. and Papp, G. and Peherstorfer, T. and Team, the ASDEX Upgrade},
    month = aug,
    year = {2024},
    pages = {106023},
}

@article{hu_plasmoid_2024,
    title = {Plasmoid drift and first wall heat deposition during {ITER} {H}-mode dual-{SPIs} in {JOREK} simulations},
    volume = {64},
    issn = {0029-5515},
    url = {https://doi.org/10.1088/1741-4326/ad53e1},
    doi = {10.1088/1741-4326/ad53e1},
    language = {en},
    number = {8},
    urldate = {2026-07-16},
    journal = {Nuclear Fusion},
    publisher = {IOP Publishing},
    author = {Hu, D. and Artola, F.J. and Nardon, E. and Lehnen, M. and Kong, M. and Bonfiglio, D. and Hoelzl, M. and Huijsmans, G.T.A. and Team, the JOREK},
    month = jun,
    year = {2024},
    pages = {086005},
}

@article{sheikh_disruption_2020,
    title = {Disruption mitigation efficiency and scaling with thermal energy fraction on {ASDEX} {Upgrade}},
    volume = {60},
    issn = {0029-5515},
    url = {https://doi.org/10.1088/1741-4326/abb425},
    doi = {10.1088/1741-4326/abb425},
    language = {en},
    number = {12},
    urldate = {2026-07-10},
    journal = {Nuclear Fusion},
    publisher = {IOP Publishing},
    author = {Sheikh, U.A. and David, P. and Ficker, O. and Bernert, M. and Brida, D. and Dibon, M. and Duval, B. and Faitsch, M. and Maraschek, M. and Papp, G. and Pautasso, G. and Sozzi, C. and team, the AUG and team, EUROfusion MST1},
    month = oct,
    year = {2020},
    pages = {126029},
}

@article{Schwarz_2023,
	doi = {10.1088/1741-4326/acf50a},
	url = {https://dx.doi.org/10.1088/1741-4326/acf50a},
	year = {2023},
	month = {sep},
	publisher = {IOP Publishing},
	volume = {63},
	number = {12},
	pages = {126016},
	author = {N. Schwarz and F.J. Artola and F. Vannini and M. Hoelzl and M. Bernert and A. Bock and T. Driessen and M. Dunne and L. Giannone and P. Heinrich and P. de Marn\'e and G. Papp and G. Pautasso and S. Gerasimov and the ASDEX Upgrade Team and JET Contributors and Team the JOREK},
	title = {The mechanism of the global vertical force reduction in disruptions mitigated by massive material injection},
	journal = {Nuclear Fusion}
}

@phdthesis{Schwarz_2024_PhD,
	author = {Schwarz, Nina},
	title = {Non-linear magneto-hydrodynamic simulations of hot and mitigated vertical displacement events and validation against experiments in the {ASDEX} {U}pgrade tokamak},
	year = {2024},
	school = {Technische Universit\"at M\"unchen},
	pages = {136},
	language = {en},
	url = {https://mediatum.ub.tum.de/1733821}
}

@article{Vallhagen_2025,
doi = {10.1088/1361-6587/ae140f},
url = {https://doi.org/10.1088/1361-6587/ae140f},
year = {2025},
month = {nov},
publisher = {IOP Publishing},
volume = {67},
number = {10},
pages = {105034},
author = {Vallhagen, O and Antonsson, L and Halldestam, P and Papp, G and Heinrich, P and Patel, A and Hoppe, M and Votta, L and {the ASDEX Upgrade Team}and {the EUROfusion Tokamak Exploitation Team}},
title = {Simulation of shattered pellet injections with plasmoid drifts in {ASDEX Upgrade} and {ITER}},
journal = {Plasma Physics and Controlled Fusion}
}

@article{veres_fast_2009,
    series = {Proceedings of the 18th {International} {Conference} on {Plasma}-{Surface} {Interactions} in {Controlled} {Fusion} {Device}},
    title = {Fast radiation dynamics during {ELMs} on {TCV}},
    volume = {390-391},
    issn = {0022-3115},
    url = {https://www.sciencedirect.com/science/article/pii/S0022311509002475},
    doi = {10.1016/j.jnucmat.2009.01.220},
    urldate = {2026-02-24},
    journal = {Journal of Nuclear Materials},
    author = {Veres, G. and Pitts, R. A. and Bencze, A. and Márki, J. and Tál, B. and Tye, R.},
    month = jun,
    year = {2009},
    pages = {835--838},
}

@article{boivin_high_1999,
    title = {High resolution bolometry on the {Alcator} {C}-{Mod} tokamak (invited)},
    volume = {70},
    issn = {0034-6748},
    url = {https://doi.org/10.1063/1.1149309},
    doi = {10.1063/1.1149309},
    number = {1},
    urldate = {2026-07-24},
    journal = {Review of Scientific Instruments},
    author = {Boivin, R. L. and Goetz, J. A. and Marmar, E. S. and Rice, J. E. and Terry, J. L.},
    month = jan,
    year = {1999},
    pages = {260--264},
}

@article{czarny_bezier_2008,
    title = {Bézier surfaces and finite elements for {MHD} simulations},
    volume = {227},
    issn = {0021-9991},
    url = {https://www.sciencedirect.com/science/article/pii/S0021999108002118},
    doi = {10.1016/j.jcp.2008.04.001},
    number = {16},
    urldate = {2026-07-26},
    journal = {Journal of Computational Physics},
    author = {Czarny, Olivier and Huysmans, Guido},
    month = aug,
    year = {2008},
    pages = {7423--7445},
}

@article{huysmans_mhd_2007,
    title = {{MHD} stability in {X}-point geometry: simulation of {ELMs}},
    volume = {47},
    issn = {0029-5515},
    shorttitle = {{MHD} stability in {X}-point geometry},
    url = {https://doi.org/10.1088/0029-5515/47/7/016},
    doi = {10.1088/0029-5515/47/7/016},
    language = {en},
    number = {7},
    urldate = {2026-07-26},
    journal = {Nuclear Fusion},
    author = {Huysmans, G.T.A. and Czarny, O.},
    month = jun,
    year = {2007},
    pages = {659},
}

@article{bandyopadhyay_mhd_2025,
    title = {{MHD}, disruptions and control physics: {Chapter} 4 of the special issue: on the path to tokamak burning plasma operation},
    volume = {65},
    issn = {0029-5515},
    shorttitle = {{MHD}, disruptions and control physics},
    url = {https://infoscience.epfl.ch/handle/20.500.14299/254382},
    doi = {10.1088/1741-4326/ade7a0},
    language = {en},
    number = {10},
    urldate = {2026-07-26},
    publisher = {IOP Publishing Ltd},
    author = {Bandyopadhyay, I. and Igochine, V. and Sauter, O. and Sabbagh, S. and Park, J. K. and Nardon, E. and Villone, F. and Maraschek, M. and Pautasso, G. and Eidietis, N. and Jardin, S. and Humphreys, D. and Dubrov, M. and Artola, F. and de Baar, M. and Bardóczi, L. and Baylor, L. and Berkery, J. and Boozer, A. and Cannas, B. and Chen, Z. and Esposito, B. and Fanni, A. and Ferraro, N. and Fitzpatrick, R. and Gerasimov, S. and Goodman, T. and Granetz, R. and Granucci, G. and Graves, J. and Gribov, Y. and Gude, A. and Hoelzl, M. and Hollmann, E. and Hu, Q. and Hu, W. and In, Y. and Isayama, A. and Isernia, N. and Jachmich, S. and Kavin, A. and Khayrutdinov, R. and Kim, G. and Kong, M. and Kudláček, O. and Lehnen, M. and Liu, Y. and Logan, N. and Lukash, V. and Maget, P. and Markovic, T. and Matsuyama, A. and Maviglia, F. and Menard, J. and Myers, C. and Orlov, D. and Pau, A. and Paz-Soldan, C. and Piron, L. and Pucella, G. and Pustovitov, V. and Rattá, G. and Rea, C. and Reimerdes, H. and Reux, C. and Roccella, R. and Rubinacci, G. and Sheikh, U. and Shiraki, D. and Sias, G. and Sieglin, B. and Sovinec, C. and Strauss, H. and Sun, Y. and Sweeney, R. and Wang, H. H. and Yang, S. and Yanovskiy, V. and Zohm, H.},
    month = oct,
    year = {2025},
    journal = {Nuclear Fusion},
}

@article{hender_chapter_2007,
    title = {Chapter 3: {MHD} stability, operational limits and disruptions},
    volume = {47},
    issn = {0029-5515, 1741-4326},
    shorttitle = {Chapter 3},
    url = {https://iopscience.iop.org/article/10.1088/0029-5515/47/6/S03},
    doi = {10.1088/0029-5515/47/6/S03},
    language = {en},
    number = {6},
    urldate = {2026-07-26},
    journal = {Nuclear Fusion},
    author = {Hender, T.C and Wesley, J.C and Bialek, J and Bondeson, A and Boozer, A.H and Buttery, R.J and Garofalo, A and Goodman, T.P and Granetz, R.S and Gribov, Y and Gruber, O and Gryaznevich, M and Giruzzi, G and Günter, S and Hayashi, N and Helander, P and Hegna, C.C and Howell, D.F and Humphreys, D.A and Huysmans, G.T.A and Hyatt, A.W and Isayama, A and Jardin, S.C and Kawano, Y and Kellman, A and Kessel, C and Koslowski, H.R and Haye, R.J. La and Lazzaro, E and Liu, Y.Q and Lukash, V and Manickam, J and Medvedev, S and Mertens, V and Mirnov, S.V and Nakamura, Y and Navratil, G and Okabayashi, M and Ozeki, T and Paccagnella, R and Pautasso, G and Porcelli, F and Pustovitov, V.D and Riccardo, V and Sato, M and Sauter, O and Schaffer, M.J and Shimada, M and Sonato, P and Strait, E.J and Sugihara, M and Takechi, M and Turnbull, A.D and Westerhof, E and Whyte, D.G and Yoshino, R and Zohm, H and Group, Disruption {And} Magnet, The Itpa Mhd},
    month = jun,
    year = {2007},
    pages = {S128--S202},
}

@techreport{parks_modeling_2016,
	title = {Modeling {Dynamic} {Fracture} of {Cryogenic} {Pellets}},
	url = {https://www.osti.gov/servlets/purl/1344852/},
	doi = {10.2172/1344852},
	language = {en},
	number = {GA--A28352, 1344852},
	urldate = {2026-08-21},
	institution = {General Atomics},
	author = {Parks, Paul},
	month = jun,
	year = {2016},
	pages = {GA--A28352, 1344852},
}

\end{document}